\documentclass[pra,superscriptaddress,amsmath,amssymb,twocolumn]{revtex4-2}
\usepackage{graphicx}
\usepackage{subfigure}
\usepackage{adjustbox}
\usepackage{bm}
\usepackage{bbm}
\usepackage{color}
\usepackage{braket}
\usepackage{standalone}
\usepackage{multirow}
\usepackage{tikz}
\usepackage{mathrsfs}
\usepackage{dsfont}
\usepackage[colorlinks,bookmarks=true,citecolor=blue,linkcolor=blue,urlcolor=blue]{hyperref}
\usepackage{cleveref}
\usepackage{comment}
\usepackage{mathtools}
\usepackage{soul}

 \Crefname{equation}{Eq.}{Eqs.}
\Crefname{figure}{Fig.}{Figs.}

\newcommand{\dd}{\mathrm{d}}
\newcommand{\vv}{\mathbf{v}}
\newcommand{\jj}{\mathbf{j}}
\newcommand{\rr}{\mathbf{r}}

\begin{document}
\title{Stability and Decay of Macrovortices in Rotating Bose Gases Beyond Mean Field}

\author{Paolo Molignini}
\email{paolo.s.molignini@jyu.fi}
\affiliation{Department of Physics and Nanoscience Center, University of Jyväskylä, P.O. Box 35 (YFL), FI-40014 University of Jyväskylä, Finland}
\affiliation{Department of Physics, Stockholm University, AlbaNova University Center, 10691 Stockholm, Sweden}

\author{M. A. Caracanhas}
\affiliation{Instituto de Física de São Carlos, Universidade de São Paulo, CP 369, 13560-970 São Carlos, SP, Brazil}

\author{V. S. Bagnato}
\affiliation{Instituto de Física de São Carlos, Universidade de São Paulo, CP 369, 13560-970 São Carlos, SP, Brazil}
\affiliation{Department of Biomedical Engineering, Texas A\&M University, College Station, Texas 77843, USA}

\author{Barnali Chakrabarti}
\affiliation{Instituto de Física de São Carlos, Universidade de São Paulo, CP 369, 13560-970 São Carlos, SP, Brazil}

\date{\today}

\begin{abstract}
We study the formation, stability, and decay of macrovortices in a rotating Bose gas confined by a Mexican-hat potential with a multiconfigurational ansatz. 
By systematically including correlations beyond the mean-field level, we map the equilibrium phase diagram and identify regimes of coexistence between vortex lattices and multiply charged central vortices. 
Quench dynamics reveals that macrovortices are robust under changes in rotation or interaction strength, sustaining clean monopole oscillations with well-separated, vorticity-dependent breathing frequencies. 
In contrast, trap quenches trigger a universal decay process mediated by vortex–phonon coupling, in which rotational energy is progressively transferred to compressible modes until the macrovortex splits into singly quantized vortices. 
Our results demonstrate that macrovortex lifetimes and decay pathways can be tuned by trap confinement, providing experimentally accessible signatures of vortex–phonon interactions and collective energy transfer in correlated quantum fluids.
\end{abstract}

\maketitle

\section{Introduction} 
The study of ultracold quantum gases, confined, guided, and excited in various trapping geometries, has emerged as a major research frontier over the past decade~\cite{Cooper_2019,RevModPhys.71.463,RevModPhys.79.235,RevModPhys.80.1215}. 
Beyond their importance in fundamental physics, the study of trap geometry, topology, and dimensionality now underpins diverse quantum technologies -- from integrated photonics and programmable atomic qubits to quantum simulators based on ultracold atoms~\cite{Pelucchi,Debnath,Bloch:2012,Yago_Malo_2024}.
 
The first experimental observation of Bose–Einstein condensation (BEC) in 1995 utilized ultracold rubidium atoms trapped in a harmonic potential~\cite{Anderson}. 
Since then, the unprecedented control achieved in trapping and manipulating ultracold atoms has made it possible to realize quantum gases in a variety of confining geometries, including box potentials~\cite{Gaunt_2013}, periodic lattice potentials~\cite{Greiner2002QuantumPT,Bloch:2005}, ring traps~\cite{PhysRevA.59.2990,PhysRevA.69.033608,PhysRevA.73.013603,Ramanathan_2011,Corman_2014,Navez_2016}, and anharmonic traps~\cite{PhysRevA.69.033608}, enabling the observation of several exotic phenomena such as quantum correlations~\cite{RevModPhys.80.885}, Josephson effects~\cite{PhysRevLett.111.205301,PhysRevLett.95.010402,Kwon_2020}, and persistent currents~\cite{PhysRevLett.99.260401,PhysRevA.86.013629}. 
Ring-shaped traps can also serve as storage devices for coherent matter waves and open pathways for novel research in microgravity environments~\cite{Carollo_2022,Aveline,Lundblad_2023}.

More recently, though, a particularly intriguing topology has been introduced by confining a quantum gas in a {\it bubble} geometry, allowing atoms to move on a closed two-dimensional surface. 
The concept of the bubble-trap potential was proposed by Garraway and Zobay in 2001~\cite{PhysRevLett.86.1195,PhysRevA.69.023605}, based on the superposition of a static inhomogeneous magnetic field and a spatially homogeneous radio-frequency (rf) field. 
The key idea is to generate an adiabatic potential such that, around the minimum at the resonance point, the surface forms a bubble -- or shell -- with the topology of a sphere~\cite{Arazo_2021,Garraway_2016}. 
In addition to providing a confinement stronger than harmonic for the condensate, the bubble trap exhibits several topological features with nontrivial consequences for the properties of quantum gases~\cite{Herve_2021}. 
Since its first experimental realization in 2003~\cite{Colombe}, there has been rapid experimental progress in controlling the geometry of rf-dressed adiabatic potentials~\cite{Morizot_2007,PhysRevA.74.023616,PhysRevA.81.031402,PhysRevA.85.053401}, leading to strong interest and applications in quantum devices~\cite{PhysRevLett.111.205301,Ryu,PhysRevLett.110.025302}.

Motivated by the ring-trap configuration, which can be effectively achieved by intersecting elliptical bubble traps, several recent theoretical works have considered an approximate realization -- a harmonic-like trap with a central potential hill -- forming a ``Mexican-hat" potential~\cite{Tomishiyo_2024,Brito_2020,Adhikari_2019}. 
To date, all studies of the Mexican-hat potential have been carried out within the mean-field (Gross–Pitaevskii) framework, where vortex configurations are obtained by varying the interaction strength and rotation frequency.
This approach describes weakly interacting bulk systems in which all particles occupy a single condensate mode.
However, modern quantum-gas platforms can reach regimes of strong interactions and ultra-fast rotation, where the single-orbital mean-field picture breaks down and quantum correlations become essential.
In finite systems of only a few tens or hundreds of atoms, enhanced quantum fluctuations further challenge the validity of the Gross–Pitaevskii equation, underscoring the need for a beyond–mean-field treatment.

To address this gap, in this work we consider a superfluid quantum gas of $N=100$ atoms confined in a Mexican-hat potential and employ the multiconfigurational time-dependent Hartree (MCTDH) method to capture the interplay between rotation and strong correlations. 
By systematically varying the interaction strength and rotation frequency, we uncover novel vortex structures ranging from superfluid and vortex-lattice phases to multicharged macrovortices and coexistence regimes of macrovortices and vortex lattices. 
A direct comparison with mean-field theory clearly establishes the stabilizing role of quantum fluctuations when the system becomes fragmented and angular momentum is redistributed among different modes.  

Beyond the stationary configurations, we further employ the MCTDH method to explore the real-time dynamics and stability of these macrovortex states under different quench protocols.
We track their evolution, their breathing motion, and decay channels, and offer practical insights into the controlled creation of long-lived, high-angular-momentum states.
This dynamical extension allows us to address how strongly correlated rotating condensates respond to sudden changes in system parameters -- a key question for understanding vortex stability and decay in nonequilibrium quantum fluids.

Our main observations are as follows:  
(i) For a rotation quench that leaves the Mexican-hat trap geometry and interactions unchanged, we find strong robustness is a generic feature for all vorticities. 
(ii) A sudden change in the interaction strength, without modifying the trap, induces breathing oscillations, as evidenced by the periodic modulation of monopole and quadrupole moments. 
The Fourier spectrum of the monopole moment exhibits a clear frequency separation among multicharged vortices, suggesting potential applications as a classical register where the breathing frequency encodes information.
(iii) Far-from-equilibrium dynamics following a trap quench is even more intriguing, leading to the decay of multicharged vortices into singly quantized vortices accompanied by the nucleation of vortex–antivortex pairs.
Regarding the decay timescales, macrovortices of charge four decay fastest, followed by those of charge three and two, while single vortices remain stable throughout the dynamics. 
A comparison with mean-field dynamics reveals that quantum fluctuations effectively shorten the lifetime, whereas tighter confinement enhances stability by reducing the number of decay channels. 
To visualize the vortex–phonon interaction, we also compute the velocity field and vorticity, which display a spiral energy flow between the central core and the outer annulus. 
Furthermore, the Helmholtz decomposition of the one-body current density into compressible (phononic) and incompressible (solenoidal) parts quantitatively captures the vortex–phonon coupling. 
Interestingly, while the incompressible current exhibits breathing motion, the compressible current displays interference patterns -- providing direct evidence of phonon-mediated instability.  

The rest of this paper is structured as follows.
In section \ref{sec:system}, we describe the system we analyze.
In section \ref{sec:methods}, we summarize the computational method and the observables we use to study the system at hand.
In section \ref{sec:results}, we present our results and discuss their significance in relation to the literature.
Finally, in section \ref{sec:conclusions}, we conclude our study and provide an outlook on potential future research directions.

\section{System and model}
\label{sec:system}
We set out to investigate the ground-state properties and dynamics of \( N \) contact-interacting bosons of mass \( m \) confined in a rotating mexican-hat potential.
We describe the system with an effective two-dimensional (2D) geometry described by the coordinates $\mathbf{r}_j=(x_j,y_j)$ for each particle.
In the rotating frame of the trap, the total Hamiltonian of this system can be written as 
\begin{equation}
\mathcal{H} = \sum_{j=1}^N \left[ T(\mathbf{r}_j) + V(\mathbf{r}_j) \right] + \sum_{j<k}^N W(\mathbf{r}_j -\mathbf{r}_k).
\label{eq:Ham1}
\end{equation}
The Hamiltonian comprises a kinetic term
\begin{equation}
T(\mathbf{r}) = -\frac{\hbar^2}{2 m}\nabla_{\mathbf{r}}^2 - \Omega L_z
\end{equation}
which includes a rotational contribution of angular speed $\Omega$ from the angular momentum
\begin{equation}
L_z = x p_y - y p_x = -i\hbar \left( x \frac{\partial}{\partial y} - y \frac{\partial}{\partial x} \right).
\end{equation}
As mentioned above, the second term in the one-body part of the Hamiltonian is a trap shaped as a Mexican-hat potential, i.e.
\begin{equation}
    V(\mathbf{r}) = \frac{1}{2} p_1 \mathbf{r}^2 + \frac{1}{2} p_2 \mathbf{r}^4 
\end{equation}
where $p_1<0$ and $p_2>0$ to guarantee the existence of local minima.
Unless stated otherwise, in this work we will employ $p_1=-0.1$, $p_2=0.05$.
The particles additionally interact via contact interactions of strength $g_0$, i.e. 
\begin{equation}
    W(\mathbf{r}_j - \mathbf{r}_k) = g_0 \delta(r_j - r_k).
\end{equation}
We remark that—strictly speaking—delta interactions are ill-defined in two dimensions in the ultraviolet (short-range) limit.
However, the physical regimes explored in this work lie safely outside the regions where such divergences become relevant.
Furthermore, delta interactions are considerably easier to handle numerically than finite-width interactions such as Gaussians.
Nevertheless, in Appendix \ref{app:Gaussian}, we verify that the physics generated by delta interactions in our setup is equivalent to that obtained using sufficiently narrow Gaussian potentials.

In our numerics, we adopt natural units corresponding to $\hbar=m=1$, and scale the Hamiltonian \eqref{eq:Ham1} by $\frac{\hbar^2}{mL^2}$, where $L$ is the natural length scale of the trap.

In this work, we first map out the ground state of the system in the rotating frame as a function of $(\Omega, g_0)$.
This results in different configuration of vortices, including states with macrovortices -- multiply quantized vortices localized at the center and surrounded by an annulus of superfluid density. 
We then probe out-of-equilibrium dynamics via three classes of sudden quenches. 
Rotation quenches change $\Omega \to \Omega'$ while keeping $V(\mathbf{r})$ and $g_0$ fixed; 
interaction quenches change $g_0 \to g_0'$ at fixed $V(\mathbf{r})$ and $\Omega$; 
and trap quenches abruptly modify $V(\mathbf{r}) \to V'(\mathbf{r})$ while other parameters are held as specified.
We will be in particular interested in quenches from the Mexican-hat configuration to a quadratic trap, with parameters changed as $(p_1, p_2)=(-0.1, 0.05) \to (0.1, 0.0)$.
This approach separates scenarios that preserve the trap landscape from those that do not, and it will prove decisive for studying macrovortex stability.


\section{Methods and observables}
\label{sec:methods}

\subsection{MCTDH-X}
To study the system described by the Hamiltonian Eq.~\eqref{eq:Ham1} in the rotating frame, we solve the corresponding many-body Schr\"{o}dinger equation
\begin{equation}
    i \hbar \partial_t \Psi(\mathbf{r}, t) = \mathcal{H} \Psi(\mathbf{r}, t)
\end{equation}
either in imaginary time (to obtain the ground state) or in real time (to probe the dynamics from an initial state) with the MultiConfigurational Time-Dependent Hartree method for indistinguishable particles (MCTDH)~\cite{Streltsov:2006, Streltsov:2007, Alon:2007, Alon:2008, Lode:2012, Lode:2016, Fasshauer:2016} implemented in the software MCTDH-X~\cite{Lin:2020, Lode:2020, Molignini:2025-SciPost, MCTDHX}.

The MCTDH method relies on an efficient expansion of the many-body wave function $\Psi(\mathbf{r}, t)$.
The expansion consists of a time-dependent superposition of permanents constructed from $M$ variationally determined single-particle functions, known as \emph{orbitals}.
Using the Frenkel-Dirac variational principle~\cite{var1, var2, TDVM81} applied to the MCTDH ansatz, we can obtain equations of motion (EOM) that govern the optimal structure of the many-body superposition, i.e. of both the coefficients and the orbital basis.
When the EOM are solved in imaginary time, we can obtain an approximation of the many-body ground state.
When the EOM are solved in real time, we instead approximate the dynamics of the many-body system.
We remark that for $M=1$, the MCTDH equations are equivalent to a mean-field treatment such as the Gross-Pitaevskii equation. 

Strictly speaking, the MCTDH approach converges to the exact solution of the many-body Schrödinger equation as $M \to \infty$.
However, even with a finite number of orbitals it is often possible to obtain \emph{numerically converged} results -- i.e. results that are identical to the true solution up to a certain numerical precision and remain unchanged upon further increasing $M$.
This is particularly true for bosonic systems in a near-condensed state, i.e. with a weak degree of fragmentation, as we study here.
In this study, we will consider up to $M=3$ orbitals.
A more detailed presentation of the MCTDH method can be found in Appendix~\ref{app:MCTDHX}.

\subsection{Density matrices}
Once the many-body wave function $\Psi(\mathbf{r}, t)$ has been obtained from the MCTDH approach, it is rather simple to calculate observables from it.
The one-body density distribution in real space is calculated as 
\begin{align} 
\rho(\mathbf{r}) &= \left< \Psi \right| \hat{\Psi}^{\dagger}(\mathbf{r}) \hat{\Psi}(\mathbf{r}) \left| \Psi \right>, 
\end{align} 
with $\Psi^{(\dagger)}(\mathbf{r})$ representing the bosonic creation (annihilation) operator at position $\mathbf{r}$, and provides information about the spatial distribution of the particles.

The one-body density distribution is the diagonal part of the more general reduced one-body density matrix (1-RDM), defined as 
\begin{align} 
\rho^{(1)}(\mathbf{r},\mathbf{r}') &= \left< \Psi \right| \hat{\Psi}^{\dagger}(\mathbf{r}) \hat{\Psi}(\mathbf{r}') \left| \Psi \right>. 
\end{align} 
From the 1-RDM, we can measure the degree of orbital occupation by performing a spectral decomposition
\begin{equation} 
\rho^{(1)}(\mathbf{r},\mathbf{r}') = \sum_i \rho_i \phi^{(\mathrm{NO}),*}_i(\mathbf{r}')\phi^{(\mathrm{NO})}_i(\mathbf{r}) 
\label{eq:RDM1} 
\end{equation} 
where the eigenvalue $\rho_i$ of the 1-RDM denotes the occupation of the $i$-th orbital (the orbitals here are ordered from most to least occupied).
This quantity measures the population of the natural orbital $\phi_i^{(\mathrm{NO})}$, i.e. the $i$-th eigenfunction of the 1-RDM.
The occupations describe the extent of wave function fragmentation in the system: when a single orbital is macroscopically occupied ($\rho_1 \approx 1$), the wave function is approximately a coherent mean-field solution, but when $\rho_1 < 1$, fragmentation into multiple orbitals occurs.

\subsection{Superfluid vortices}
In general, the natural orbitals $\phi_i^{(\mathrm{NO})}(\mathbf{r})$ are complex-valued functions that can be expressed in polar form as 
\begin{equation}
    \phi_i^{(\mathrm{NO})}(\mathbf{r}) = |\phi_i^{(\mathrm{NO})}(\mathbf{r})| e^{i \theta_i(\mathbf{r})},
\end{equation}
where $|\phi_i^{(\mathrm{NO})}(\mathbf{r})|$ and $\theta_i(\mathbf{r})$ denote, respectively, the local amplitude and phase of the orbital. 
The amplitude $|\phi_i^{(\mathrm{NO})}(\mathbf{r})|^2$ gives the probability density distribution associated with the orbital, while the spatial variation of the phase $\theta_i(\mathbf{r})$ carries information about the local flow of the quantum probability current. 
A \emph{vortex} corresponds to a point (in two dimensions) where the amplitude of the orbital vanishes and the phase becomes undefined.
Due to the single-valuedness of the many-body wave function, around such a point the phase exhibits a \emph{quantized winding}, meaning that the integral of the phase gradient along a closed path $\mathcal{C}$ encircling the vortex satisfies
\begin{equation}
    \oint_\mathcal{C} \nabla \theta_i(\mathbf{r}) \cdot d\mathbf{r} = 2\pi \ell,
\end{equation}
with $\ell \in \mathbb{Z}$ the \emph{vorticity} of the vortex. 
In numerical representations, these phase windings appear as \emph{phase discontinuities} that occur along lines connecting regions of low density, i.e. sudden jumps of $2\pi$ in the argument of $\phi_i^{(\mathrm{NO})}$.
Therefore, identifying phase discontinuities in the most occupied natural orbital provides a clear signature of the presence and position of vortices in the system.

\subsection{Collective motion and vortex decay}
After a quench, the system exhibits collective oscillations that can be characterized in terms of low-lying excitation modes. 
In the annular geometry of the Mexican-hat trap, two dominant types of motion emerge: monopole and quadrupole oscillations. 
In our simulations, rotation and interaction quenches primarily excite the monopole mode, leading to nearly isotropic breathing oscillations of the annulus, while trap quenches introduce a stronger coupling to quadrupole distortions. 

The monopole or breathing mode corresponds to an overall radial expansion and contraction of the annulus surrounding the macrovortex, preserving its circular symmetry.
To quantify the monopole oscillations, we monitor
\begin{equation}
    \langle x^2+y^2\rangle(t) = \int \dd x \dd y \, (x^2 + y^2) \rho(x,y;t).
\end{equation}
The dominant frequencies of the monopole moment depend sensitively on the vorticity $\ell$ of the macrovortex and can be extracted from the time trace $\langle x^2 + y^2 \rangle (t)$ by performing a Fourier analysis.

In contrast, the quadrupole mode involves angular distortions of the density profile, where opposite sectors of the ring alternately stretch and compress, breaking the azimuthal symmetry of the cloud while conserving its area.
This type of motion is associated with macrovortex decay and its oscillations occur at distinct frequencies from those of the monopole.
The quadrupole mode can be measured from
\begin{equation}
    \langle x^2 - y^2\rangle(t) = \int \dd x \dd y \, (x^2 - y^2) \rho(x,y;t).
\end{equation}
In addition to this component, we also monitor
\begin{equation}
    \langle xy \rangle(t) = \int \dd x \dd y \, x y\, \rho(x,y;t),
\end{equation}
which captures quadrupole deformations rotated by $45^\circ$ with respect to the $x$ and $y$ axes. 
Together, $\langle x^2 - y^2\rangle$ and $\langle xy\rangle$ form the two orthogonal components of the quadrupole tensor, allowing us to distinguish pure elongations from shearing or rotated distortions of the density profile. 

Besides exciting collective motion, quench dynamics can also induce macrovortex decay.
To measure this, we use an operational definition of the macrovortex lifetime $t_{\mathrm{dec}}$ as the earliest time when the distance between the central vortex singularities grows beyond a fixed threshold, which we set to be $r_{\mathrm{thr}}=0.05$.
This criterion agrees with visual core collapse and phase unwinding in the figures.

\subsection{Helmholtz decomposition}
To further describe the vortex dynamics, we consider quantities derived from the spectral decomposition of the 1-RDM, for example the corresponding one-body density
\begin{equation}
    n(\rr,t)=\sum_k n_k|\phi_k|^2
\end{equation} 
and the one-body current
\begin{equation}
    \jj(\rr,t)=\frac{\hbar}{m}\sum_k n_k(t)\,\mathrm{Im}\!\left[\phi_k^*(\rr,t)\nabla \phi_k(\rr,t)\right].
\end{equation}
From the ratio of the two, we can obtain the velocity field $\vv=\jj/n$, which describes the instantaneous particle flow in the 2D space. 
The scalar vorticity $\omega_z(\rr,t) = [\nabla\times\vv]_z$ instead tracks the circulation and the integrity of the vortex cores. 

To further diagnose the nature of the flow, we perform a Helmholtz decomposition~\cite{Jackson:1998} of the current into an incompressible and a compressible contribution,
\begin{equation}
    \jj=\jj^{(i)}+\jj^{(c)},\qquad \nabla\!\cdot\!\jj^{(i)}=0,\quad \nabla\!\times\!\jj^{(c)}=\mathbf{0}.
\end{equation}
The incompressible (or solenoidal) component captures vorticity and rotational flow around vortices, while the compressible part is associated with density waves and phonon-like excitations~\cite{Bradley:2012, White:2014}.
Their relative strengths quantify how the energy of the system is partitioned
between rotational and longitudinal modes.

\subsection{Kinetic energy decomposition}
By further considering the kinetic energy stemming from each Helmholtz component, we can indirectly measure the nature and magnitude of the excitations in the system.
The total kinetic energy density can be expressed as
\begin{equation}
    \mathcal{E}_{\text{kin}}(\mathbf{r}) = 
        \frac{m}{2} \frac{|\mathbf{j}(\mathbf{r})|^2}{n(\mathbf{r})}.
\end{equation}
Integrating this quantity over space for each Helmholtz component yields the incompressible and compressible kinetic energies:
\begin{equation}
    E_i = \frac{m}{2} \int d^2r \,
        \frac{|\mathbf{j}^{(i)}(\mathbf{r})|^2}{n(\mathbf{r})}, \,
    E_c = \frac{m}{2} \int d^2r \,
        \frac{|\mathbf{j}^{(c)}(\mathbf{r})|^2}{n(\mathbf{r})}.
\end{equation}
$E_i$ reflects solenoidal, vortex-dominated motion, whereas $E_c$ captures compressible, phonon-like flow.

We further define the ratio
\begin{equation}
    R_c = \frac{E_c}{E_i + E_c},
\end{equation}
which provides a compact measure of the relative weight of compressible excitations. 
A low $R_c$ indicates a flow dominated by vortices and rotational motion, while a higher $R_c$ reveals increasing contributions from sound-like excitations or density modulations.
At long times, the system typically reaches a mixed regime in which both vortical and phononic excitations coexist and share comparable kinetic energy.
This results in a saturation value $R_c \simeq 0.5$, where rotational and compressional motions are approximately in equipartition, as observed in two-dimensional quantum turbulence (e.g. Nore et al., Phys. Fluids 9, 2644 (1997); Kobayashi and Tsubota, PRL 94, 065302 (2005)).

Finally, to monitor the energy flow between the incompressible and compressible components of the current as a probe of vortex-phonon conversion, we compute the time derivatives
\begin{equation}
    \dot{E}_i(t) = \frac{dE_i}{dt}, \qquad
    \dot{E}_c(t) = \frac{dE_c}{dt}.
\end{equation}
A correlated decrease in $E_i$ and increase in $E_c$ signals the transfer of rotational kinetic energy into sound-like modes -- a hallmark of vortex-phonon interaction.


\section{Results}
\label{sec:results}

\begin{figure}
    \centering
    \includegraphics[width=\columnwidth]{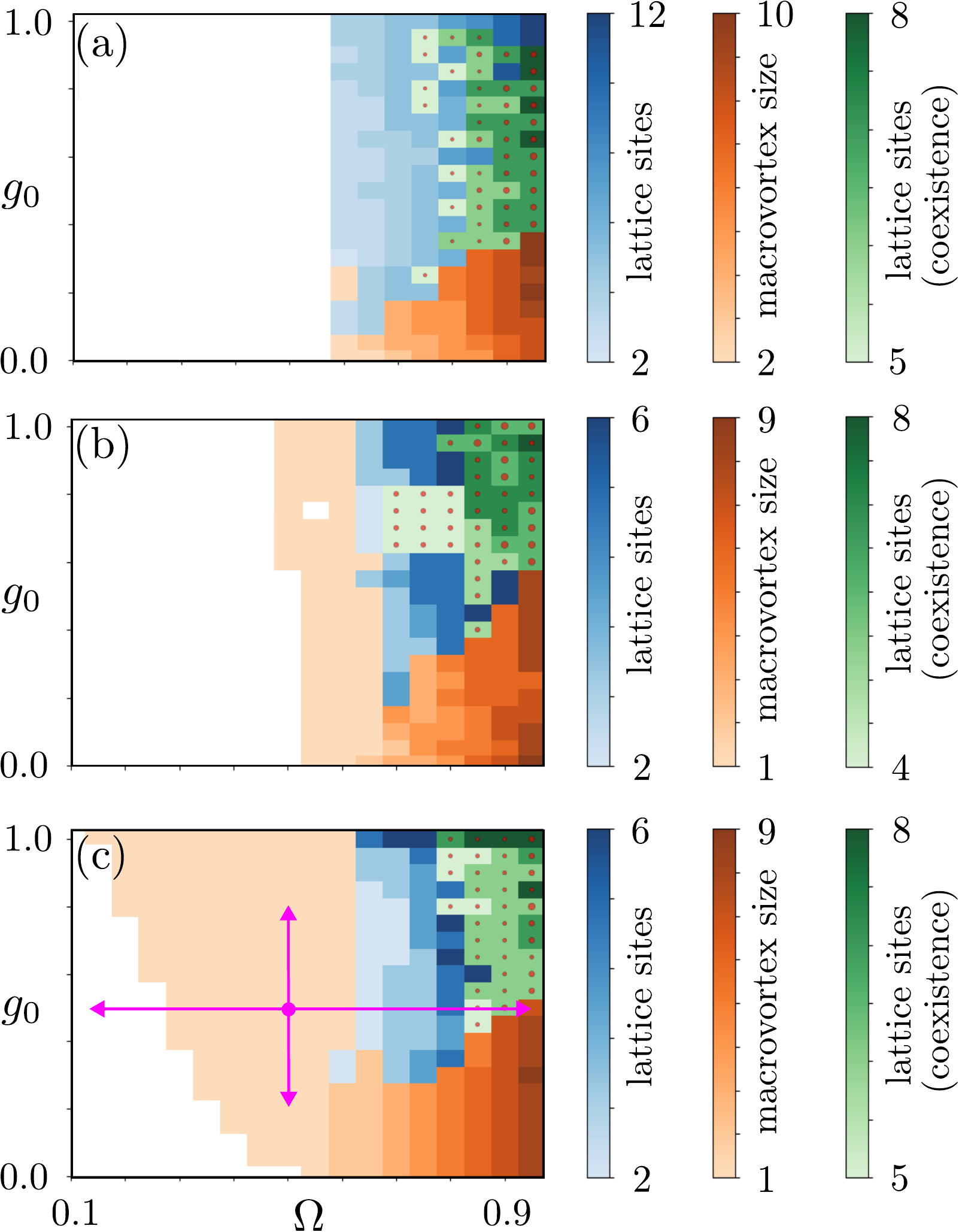}
    \caption{\textbf{Phase diagram for $N=100$ rotating bosons in a Mexican-hat potential.} The diagram is plotted as a function of angular velocity $\Omega$ and interaction strength $g_0$, plotted for (a) $M=1$ orbital (mean-field result), (b) $M=2$ orbitals, and (c) $M=3$ orbitals. The different color schemes indicate the three possible phases: vortex lattice only (blue colorbar), central macrovortex only (orange colorbar), and coexistence of the two (green colorbar). For the region where the central macrovortex coexists with a surrounding vortex lattice, the red dots additionally mark the macrovortex size.
    The pink arrows in panel (c) indicate the approximate extent of sudden quenches used to assess vortex stability, here only showed for a single-vortex state.}
    \label{fig:PD}
\end{figure}

\subsection{Ground state}
The equilibrium phases of the rotating Bose gas in the Mexican-hat potential are summarized in Fig.~\ref{fig:PD} as a function of the rotation frequency $\Omega$ and the interaction strength $g_0$ for different numbers of orbitals $M$.
The panels illustrate how including correlations progressively modifies the structure of the phase diagram and the stability ranges of the various vortex states.
However, in each case we can distinguish four main behaviors.
At low values of $g_0$ and $\Omega$, the system always arranges as a single superfluid droplet with no vortices (white region).
As $\Omega$ increases for weak interactions, a phase with a single macrovortex in the center appears (orange color scheme).
As interactions are increased, the centrifugal energy becomes large enough to nucleate individual vortices which arrange themselves in a lattice (blue color scheme).
When both $g_0$ and $\Omega$ are large enough, the system transitions into a coexistence of macrovortices and vortex lattice (green color scheme with red dots).

\begin{figure}
    \centering
    \includegraphics[width=\columnwidth]{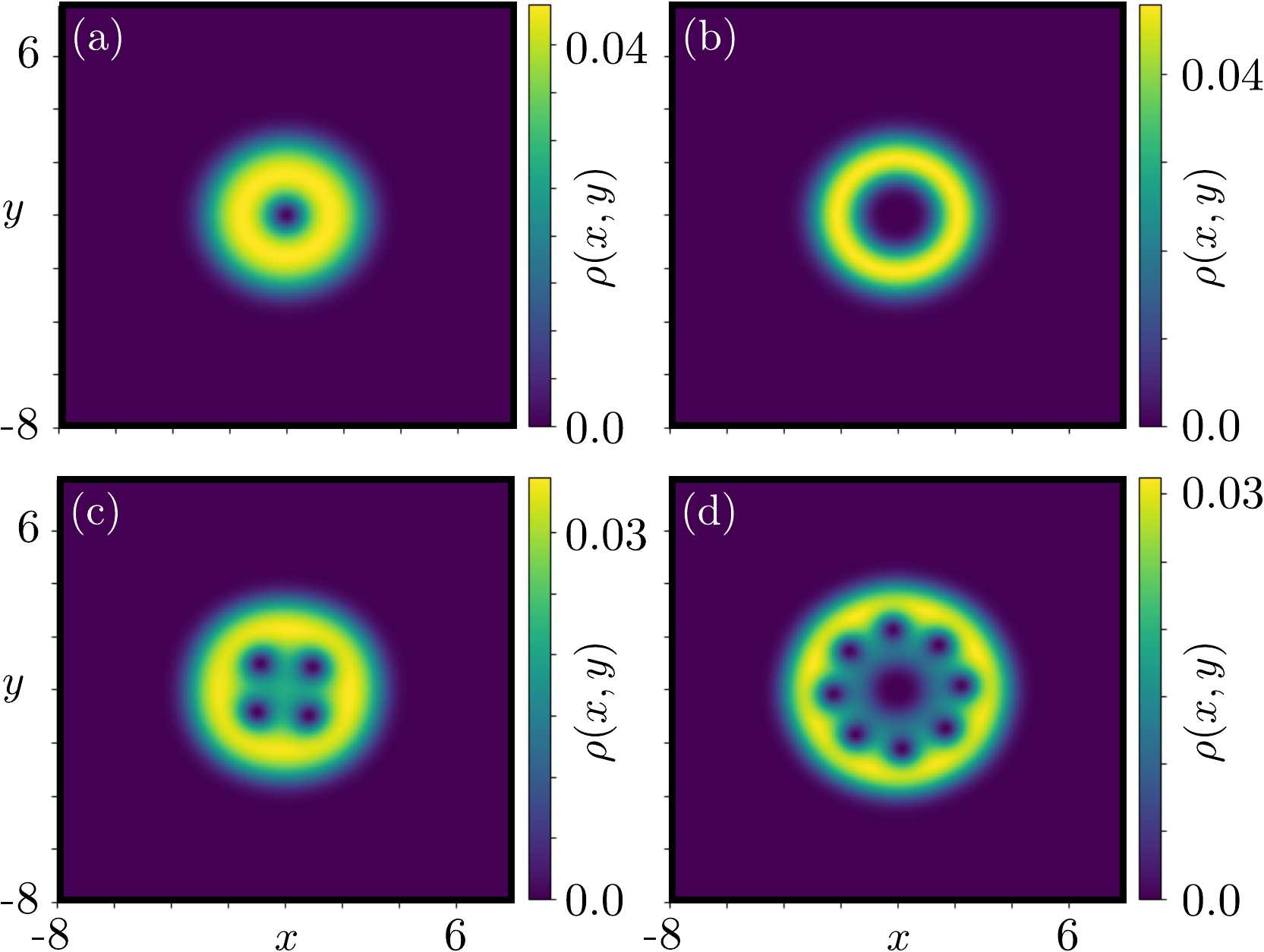}
    \caption{\textbf{Representative density distributions for the different ground states encountered in the rotating Mexican hat potential.} (a) single vortex ($g_0=0.5$, $\Omega=0.5$), (b) vortex lattice ($g_0=0.9$, $\Omega=0.75$), (c) macrovortex -- here with vorticity 3 ($g_0=0.2$, $\Omega=0.7$), (c) coexistence of macrovortex -- here with vorticity 2 -- and vortex lattice ($g_0=0.95$, $\Omega=0.95$). All the densities were obtained for $N=100$ bosons and $M=3$ orbitals. }
    \label{fig:moments-GS}
\end{figure}

Representative density profiles of the different phases are shown in Fig.~\ref{fig:moments-GS} for $M=3$.
Panel (a) displays the simplest nontrivial configuration, a single vortex in the center of the trap. 
A macrovortex containing multiple units of quantized angular momentum and a resulting annular density is displayed in panel (b).
Panel (c) shows an example of a vortex lattices with four vortex cores.
Finally, the coexistence regime is shown in panel (d), with a macrovortex of charge 2 surrounded by 8 single vortices.

We now turn to the effect of quantum fluctuations, as captured by the inclusion of additional orbitals, on the overall structure of the phase diagram. 

At the mean-field level ($M=1$), the system displays a  sharp transition between a featureless superfluid and a well-formed vortex lattice, occurring around $\Omega \simeq 0.6$. 
Interestingly, at this level of description and within the parameter resolution probed, we do not observe an intermediate state corresponding to a single, isolated vortex. 
The crossover between the vortex lattice and the macrovortex regime appears already at very weak interactions, around $g_0 \simeq 0.1$, indicating that at mean field the macrovortex is stabilized only in a narrow window of low coupling and moderate rotation.

When correlations are partially included ($M=2$), the boundaries separating these phases begin to soften. 
The onset of vortices shifts toward slightly lower rotation frequencies ($\Omega \sim 0.5$), and the coexistence region between lattice and macrovortex configurations becomes broader and more continuous. 
The inclusion of a second orbital allows the system to partially fragment and to redistribute angular momentum among different modes, leading to smoother transitions and reduced sensitivity to the precise values of $g_0$ and $\Omega$.

A more pronounced restructuring occurs when three orbitals are included ($M=3$), as shown in panel (c) of Fig.~\ref{fig:PD}. 
The superfluid region shrinks considerably, and isolated single-vortex states emerge already at much lower rotation frequencies.
For instance, at $g_0=1.0$ we find vortices forming at $\Omega \approx 0.15$. 
At the same time, the onset of the vortex-lattice phase is pushed toward higher interaction strengths, around $g_0 \sim 0.35$. 
Moreover, we also observe smoother transitions between vortex states with different vorticity, specially for the macrovortex region.
The coexistence region becomes less patchy and more structured.

This behavior signals a stabilizing role of quantum fluctuations: by allowing correlations between orbitals, the system can better accommodate the rotational flow without fragmenting into a full lattice, thereby extending the stability of macrovortex states to stronger interactions and slower rotation. 
In other words, fluctuations effectively smooth out the energy landscape and raise the barrier between competing vortex configurations.
This signals that in experimental settings where quantum fluctuations are present, multi-charged vortex configurations should be more stable than what mean-field theory predicts.

\subsection{Rotation and interaction quenches}

\begin{figure*}[t!]
    \centering
    \includegraphics[width=\textwidth]{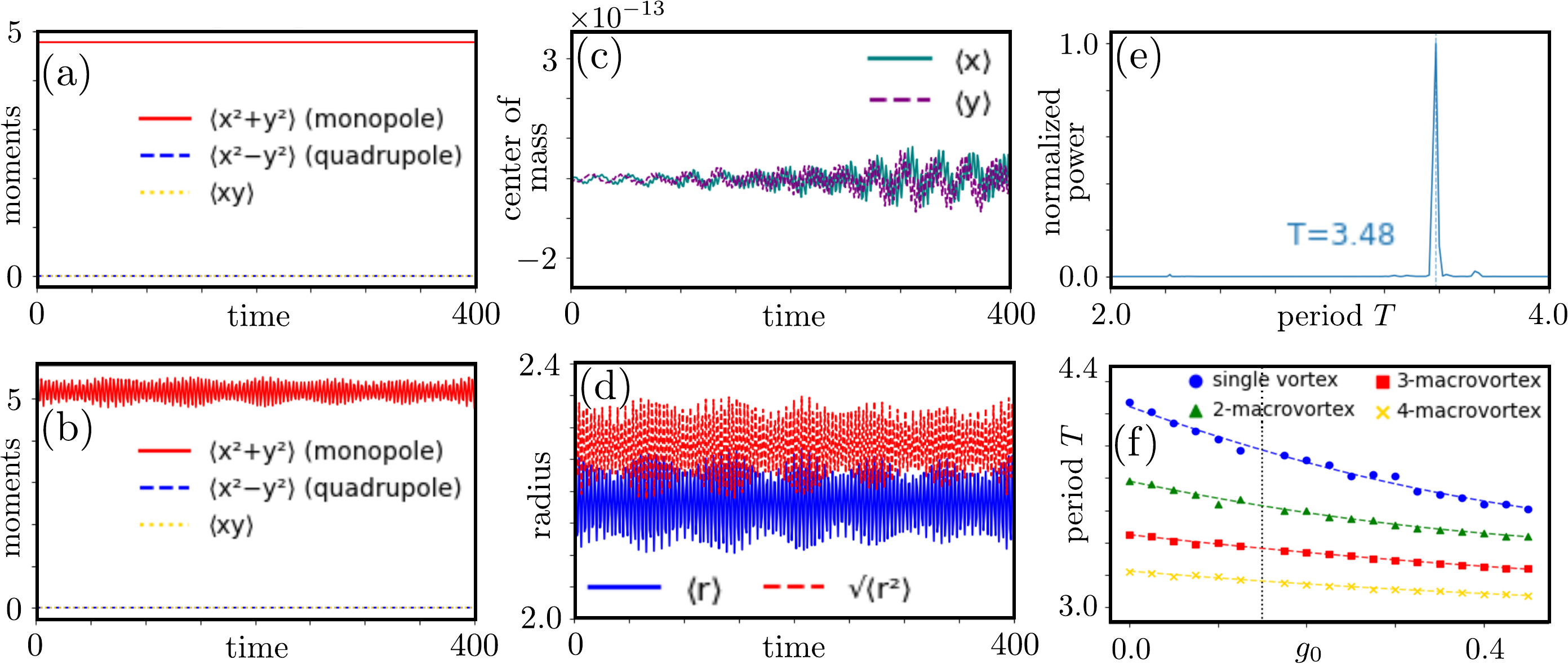}
    \caption{\textbf{Macrovortex stability in rotation and interaction quench.}
    (a) Dynamics of monopole and quadrupole moments for a representative rotation quench with $g_0 = 0.15$, $\Omega = 0.65 \to 0.75$.
    (b) Dynamics of monopole and quadrupole moments for a representative interaction quench with $g_0 = 0.15 \to 0.35$, $\Omega = 0.65$.
    (c) Center of mass dynamics for the interaction quench in (b).
    (d) Dynamics of the mean cloud radius for the interaction quench in (b).
    (e) Fourier transform of the monopole time series in (b).
    (f) Breathing mode oscillation period as a function of the strength of the quenched interaction for initial states exhibiting macrovortices with different vorticity. The vertical dotted line at $g_0=0.15$ indicates the initial interaction strength. The dashed lines are quadratic fits.
    }
    \label{fig:moments}
\end{figure*}

We now analyze the dynamical response of the macrovortex states under two types of quenches: rotation and interaction quenches. 
The results are summarized in Fig.~\ref{fig:moments}.

For rotation quenches, shown in panel (a), we find that the system remains remarkably stable. 
The density profile and all low-order moments remain essentially constant in time, and even the monopole moment $\langle x^2 + y^2 \rangle$ shows no appreciable oscillation. 
The absence of any collective excitation indicates that the macrovortex state is an eigenstate of the rotating frame Hamiltonian for the new rotation frequency as long as the trapping potential remains unchanged. 
This behavior is generic for all vorticities we probed: macrovortices remain completely robust under moderate changes of the rotation frequency when the shape of the Mexican-hat potential is preserved. 
This stability highlights their potential as persistent, rotation-insensitive flow configurations.

A different scenario emerges for interaction quenches, illustrated in panels (b)–(f). 
Following a sudden change in the interaction strength $g_0$, the annular condensate undergoes pronounced radial breathing oscillations. 
These are reflected as clear periodic modulations in the monopole moment $\langle x^2 + y^2 \rangle(t)$ [panel (b)], while both quadrupole components, $\langle x^2 - y^2 \rangle$ and $\langle xy \rangle$, remain essentially constant. 
The dynamics are thus purely isotropic: the ring expands and contracts without any angular distortion. 
The motion of the center of mass [panel (c)] remains negligible, confirming that the oscillations correspond to genuine breathing rather than drift. 
This breathing behavior is also evident in the mean cloud radius and its variance [panel (d)], which oscillate as the annular density pulsates in and out.

The Fourier spectrum of the monopole signal [panel (e)] reveals a single dominant frequency, sometimes accompanied by weak beating patterns, indicating the presence of a well-defined collective mode slightly modulated by interaction-induced anharmonicities. 
The extracted breathing frequency depends on the post-quench interaction quadratically, but with a very small quadratic coefficient that make the overall behavior nearly linear, as shown in panel (f). 
Importantly, the frequencies associated with macrovortices of different vorticity $\ell$ are clearly separated. 
This distinct spectral spacing implies that the breathing response of the condensate can be used to unambiguously identify the vorticity of a macrovortex simply by measuring its oscillation frequency.

The clear frequency separation between macrovortices of different charge also points to a potential application beyond simple state identification. 
In principle, each macrovortex could serve as a classical register, with information encoded in its characteristic breathing frequency. 
The fact that these modes remain spectrally well isolated means that multiple vorticity states could be addressed independently within the same system without cross-talk. Extending this idea further, if one could engineer coherent superpositions of states with opposite circulation, such as $\ket{+m} + \ket{-m}$, and dynamically control the interconversion between different vorticities, the system could serve as a building block for a fully quantum register based on rotational degrees of freedom. 
Realizing such coherent vortex superpositions, however, would require fine control over the trap and interaction landscape and lies beyond the scope of the present study.

\subsection{Trap quenches and vortex decay}

\begin{figure}[h!]
    \centering
    \includegraphics[width=0.9\columnwidth]{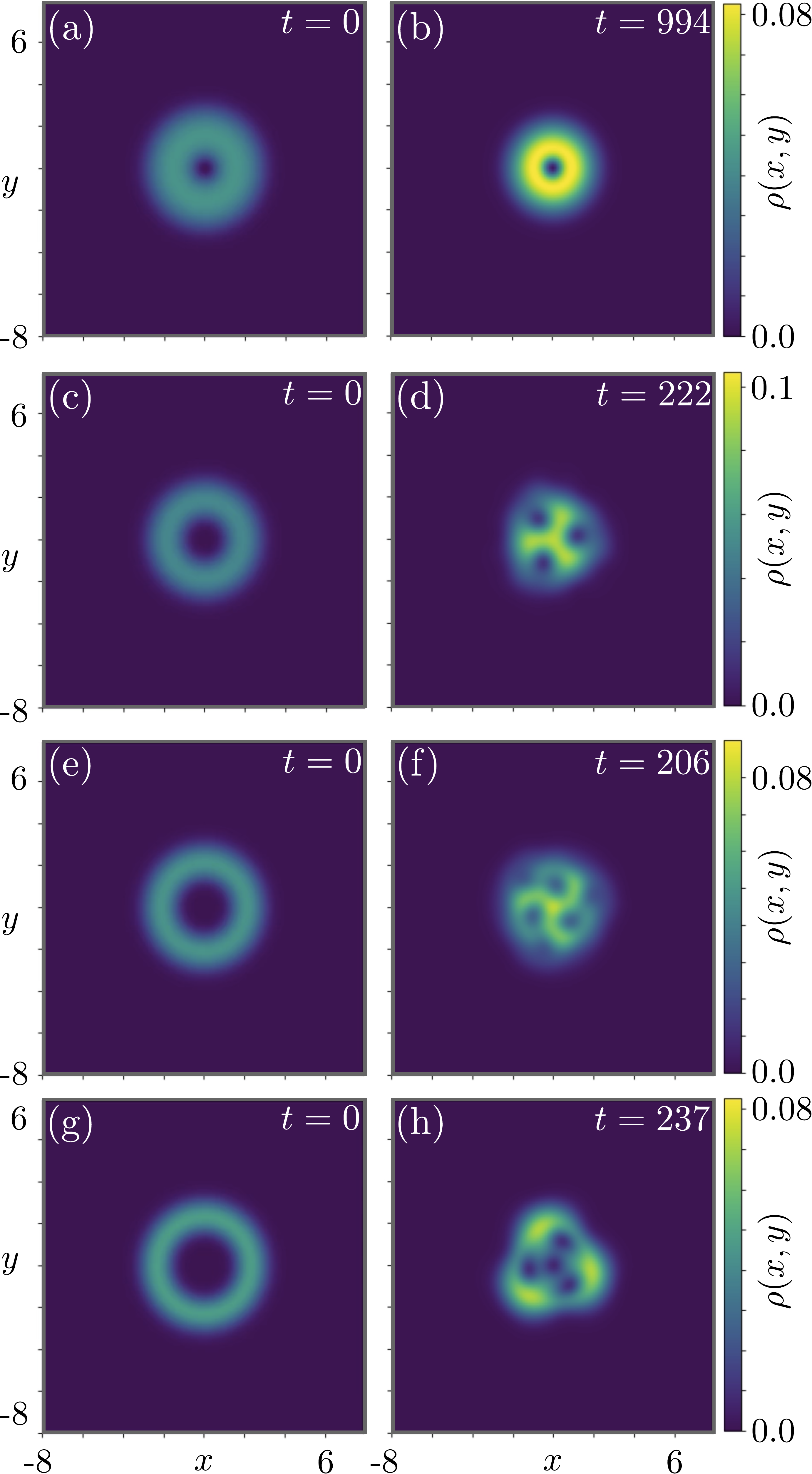}
    \caption{\textbf{Density behavior after a trap quench.} 
    The panels show the initial density (left panels) and the density after decay if any (right panels) for states with initial vortices with vorticity 1 through 4. 
    The plots were obtained for $N=100$ bosons in $M=3$ orbitals with quenched trap parameters $(p_1, p_2)=(-0.1, 0.05) \to (0.1, 0.0)$. 
    The parameters for the initial states are 
    (a)-(b) $g_0=0.5$, $\Omega=0.5$ (single vortex),      
    (c)-(d) $g_0=0.2$, $\Omega=0.65$ (2-macrovortex),  
    (e)-(f) $g_0=0.2$, $\Omega=0.7$ (3-macrovortex),  
    (g)-(h) $g_0=0.15$, $\Omega=0.75$ (4-macrovortex).     
    }
    \label{fig:density-trap-quench}
\end{figure}

\begin{figure}[h!]
    \centering
    \includegraphics[width=0.9\columnwidth]{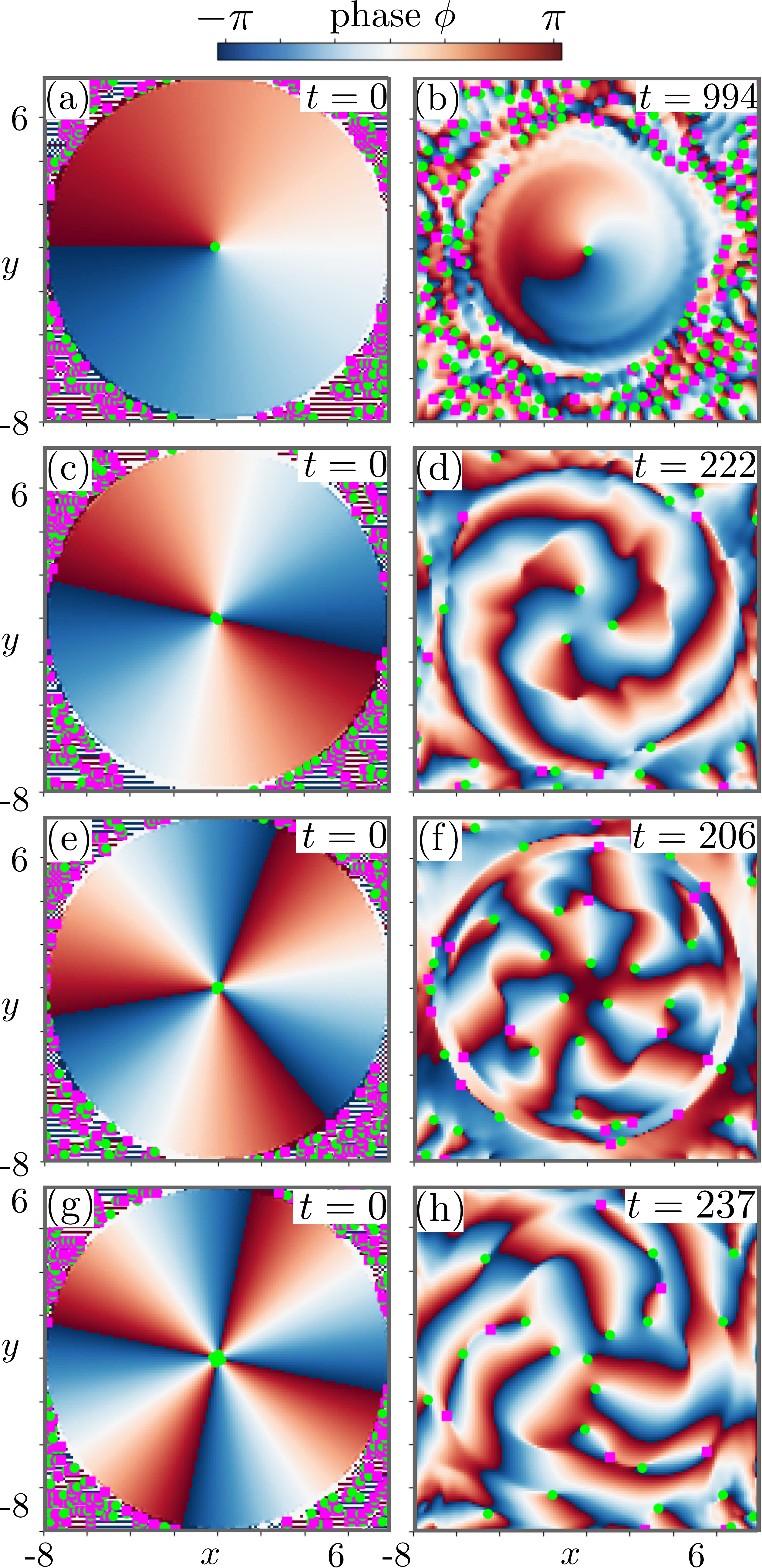}
    \caption{\textbf{Phase behavior after a trap quench.} 
    The panels show the phase of the largest natural orbital at time $t=0$ (left panels) and after decay if any (right panels), for states with initial vortices with vorticity 1 through 4. 
    The plots were obtained for $N=100$ bosons in $M=3$ orbitals with quenched trap parameters $(p_1, p_2)=(-0.1, 0.05) \to (0.1, 0.0)$. 
    The parameters for the initial states are 
    (a)-(b) $g_0=0.5$, $\Omega=0.5$ (single vortex),      
    (c)-(d) $g_0=0.2$, $\Omega=0.65$ (2-macrovortex),  
    (e)-(f) $g_0=0.2$, $\Omega=0.7$ (3-macrovortex),  
    (g)-(h) $g_0=0.15$, $\Omega=0.75$ (4-macrovortex).  
    The green dots (fuchsia squares) are a guide to the eye for vortices (antivortices). Note that at $t=0$, a number of artifact vortices appears in boundary regions where the many-body amplitude is vanishingly small.
    }
    \label{fig:phase-trap-quench}
\end{figure}

We now turn to the case of trap quenches, where the shape of the external potential is suddenly modified while keeping all other parameters fixed. 
This perturbation injects a large amount of energy into the system and drives it far from equilibrium. 
For brevity, we discuss results for macrovortices with vorticities $m=2$, $3$, and $4$, and compare them to the dynamics of a single vortex when relevant. 
The density and phase dynamics of representative cases are illustrated in Figs.~\ref{fig:density-trap-quench} and~\ref{fig:phase-trap-quench}.

As shown in Figs.~\ref{fig:density-trap-quench} and~\ref{fig:phase-trap-quench}, all macrovortex states ultimately decay into configurations of singly quantized vortices.
The decay proceeds differently depending on the initial vorticity. 
For the 2-macrovortex, the central hole first splits into two distinct vortices; one of them is subsequently absorbed from the outer region to form a stable triangular hole configuration, which reflects the local preference of the system for an Abrikosov-type triangular arrangement.
The 3-macrovortex decays directly into the triangular Abrikosov configuration.
The 4-macrovortex instead decays into a threefold-symmetric configuration with an additional vortex pinned at the trap center, a geometry that appears favored by the quadratic confinement after the quench. 
In every case, the decay is clearly visible both in the density and in the corresponding phase maps, where the initially smooth phase winding breaks up into multiple localized singularities.

\begin{figure}[t!]
    \centering
    \includegraphics[width=0.8\columnwidth]{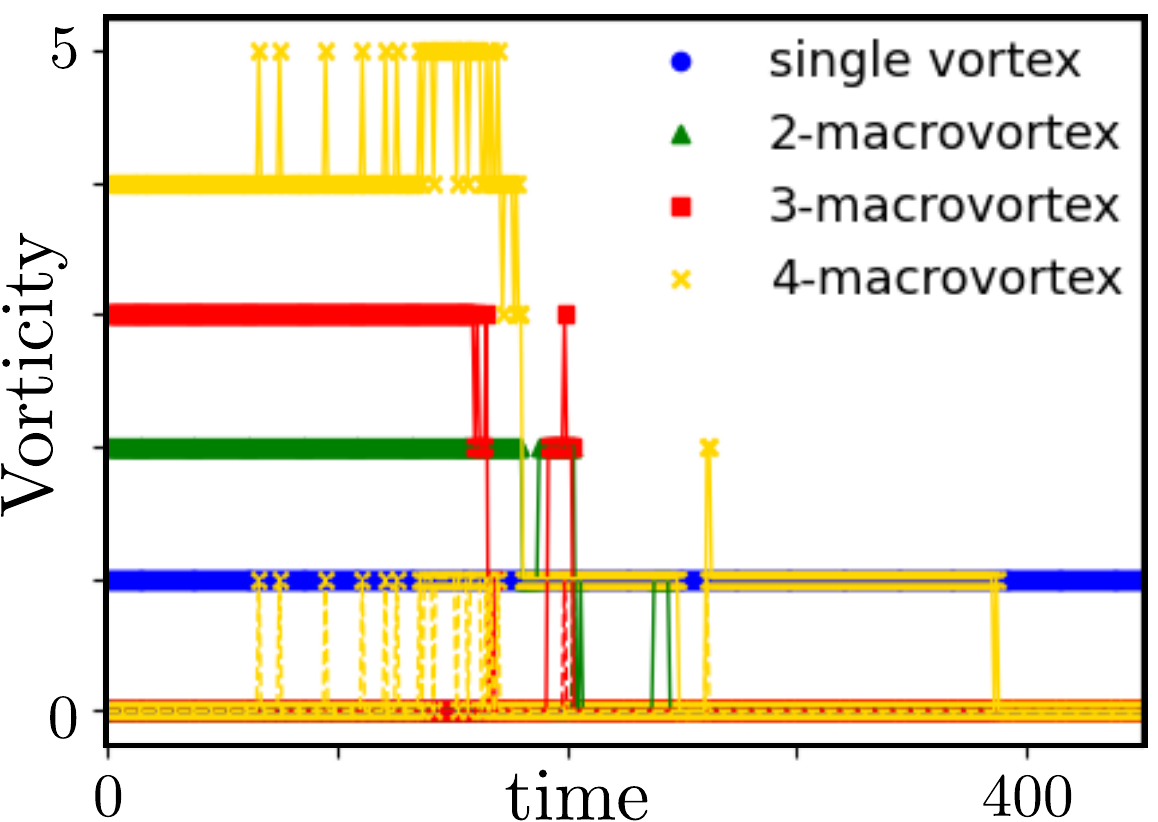}
    \caption{\textbf{Central vortex count after a trap quench.} 
    The plot shows the number of central vortex cores (vorticity) as a function of time after the trap quench for an initial state with vorticity 1 to 4.
    The plots were obtained for $N=100$ bosons and $M=3$ orbitals. The parameters for the initial states are as in Fig.~\ref{fig:decay}
    All initial states are prepared in the same Mexican-hat potential with $p_1=-0.1$, $p_2=0.05$.
    }
    \label{fig:vorticity}
\end{figure}

The evolution of the central vorticity, shown in Fig.~\ref{fig:vorticity}, provides a visualization of the  speed of the decay process. 
In all cases, the macrovortex retains its charge initially, but a sharp transition to a fragmented vortex distribution eventually occurs.
In absolute timescales, the 4-macrovortex decays first, followed by the 3- and 2-macrovortices.
The single-vortex configuration, instead, remains stable throughout the simulation window. 
This hierarchy of decay times is consistent with the intuitive notion that multiply charged vortices are more energetically costly and thus more prone to splitting once the trapping potential is perturbed.

It is also interesting to consider the role of other vortices generated during time evolution.
Immediately after the quench, a dense cloud of vortex–antivortex pairs is nucleated throughout the condensate [see appendix ~\ref{app:vortex-generation} for more details]. 
These additional vortices, however, tend to be confined to the periphery of the macrovortex core and do not interfere with it.
Even when vortex-antivortex pairs are detected within the central region where the macrovortex resides (this is particularly pronounced for the 4-macrovortex as shown in the yellow curve of Fig.~\ref{fig:vorticity}), they do not lead to an immediate breakdown of the macrovortex.
This seems to indicate that the decay mechanism is not strongly connected with vortex-vortex interactions, and instead follow a different channel. 
We will explore this question more in detail in the next section.

\begin{figure}[t!]
    \centering
    \includegraphics[width=0.8\columnwidth]{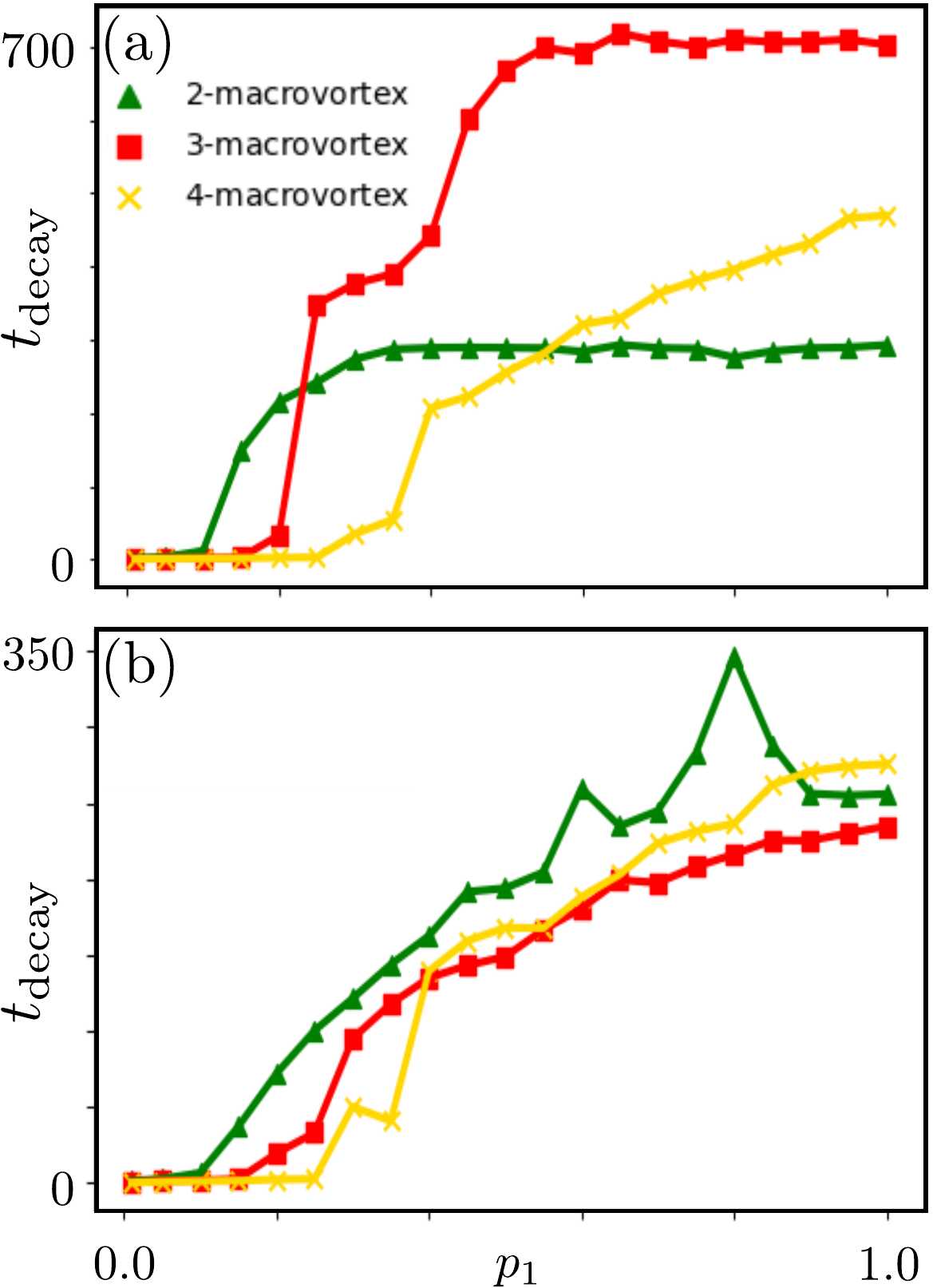}
    \caption{\textbf{Macrovortex decay time after a trap quench.} 
    The decay time is plotted as a function of the quenched trap parameter $p_1$.
    The plots were obtained for $N=100$ bosons and (a) $M=1$ (mean-field), (b) $M=3$ orbitals. The parameters for the initial states are 
    (a) $g_0=0.2$, $\Omega=0.65$ (2-macrovortex),  
    $g_0=0.2$, $\Omega=0.7$ (3-macrovortex),  
    $g_0=0.15$, $\Omega=0.75$ (4-macrovortex) and
    (b) $g_0=0.05$, $\Omega=0.6$ (2-macrovortex),  
    $g_0=0.05$, $\Omega=0.65$ (3-macrovortex),  
    $g_0=0.1$, $\Omega=0.7$ (4-macrovortex).
    All initial states are prepared in the same Mexican-hat potential with $p_1=-0.1$, $p_2=0.05$.
    }
    \label{fig:decay}
\end{figure}

To quantify macrovortex stability across different parameter choices, we calculate the decay time $t_{\mathrm{decay}}$ defined in Sec.~\ref{sec:methods}. 
The dependence of $t_{\mathrm{decay}}$ on the (quenched) trap parameter $p_1$ is shown in Fig.~\ref{fig:decay} for both $M=1$ (mean-field) and $M=3$ (correlated) simulations. 
We find that all macrovortices eventually decay, but the characteristic timescales depend strongly on both the vorticity and the degree of correlation. 
As already observed in Fig.~\ref{fig:vorticity}, macrovortices with higher winding numbers are systematically less stable and decay faster. 
Moreover, the inclusion of quantum fluctuations at $M=3$ significantly shortens the lifetime compared to mean-field dynamics, reflecting the additional decay channels enabled by correlations between orbitals.

\subsection{Decay mechanism}
In the previous section, we have seen that vortex-vortex interaction is not a likely culprit for macrovortex decay.
In this section, we examine more in detail another possible destabilization channel, namely vortex-phonon interactions.

\begin{figure}[t!]
    \centering
    \includegraphics[width=0.85\columnwidth]{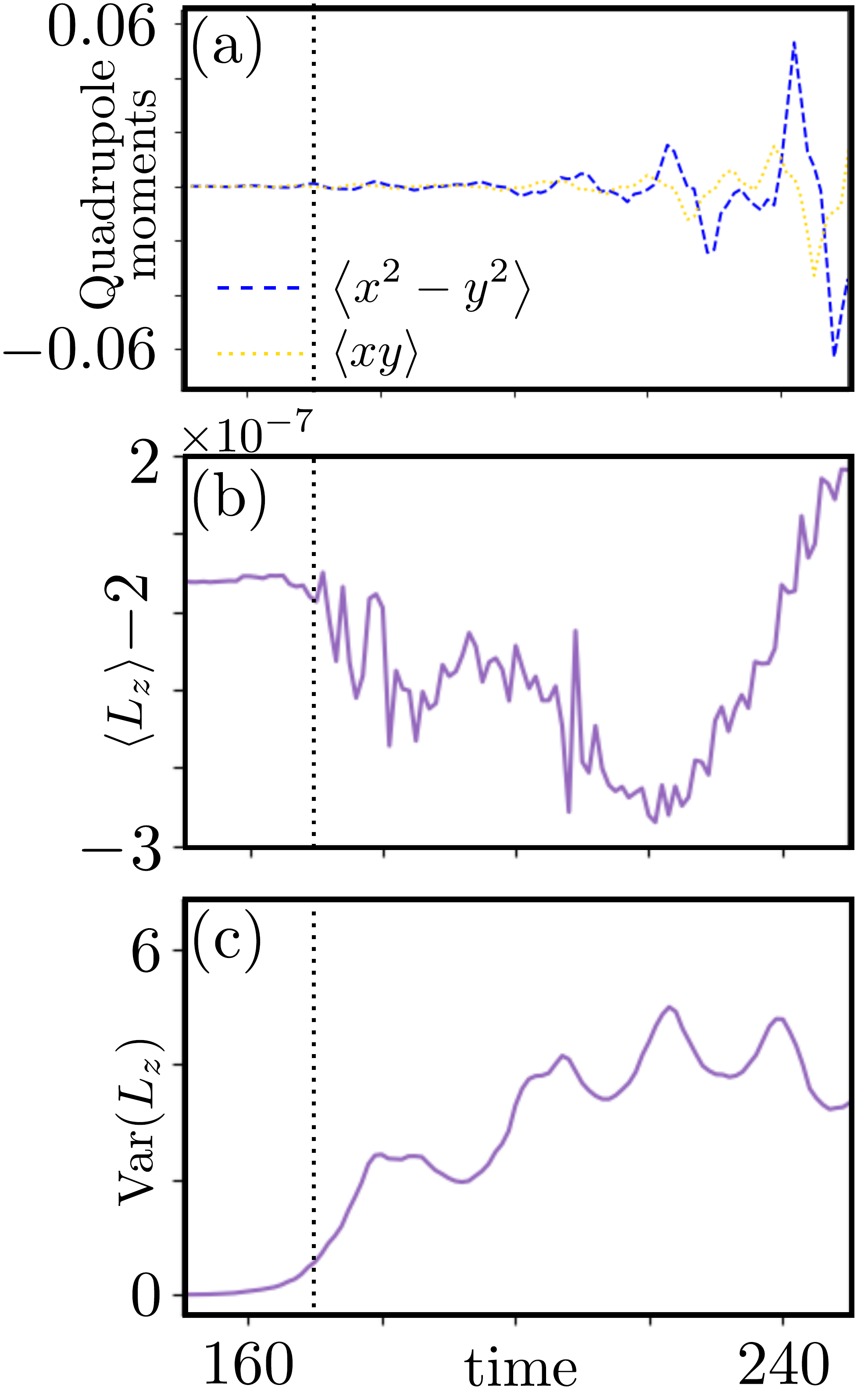}
    \caption{\textbf{Macrovortex decay signatures in quadrupole moments and angular momentum.} 
    Behavior of (a) the quadrupole moments, (b) the deviation of the mean angular momentum $\left< L_z \right>$ from vorticity 2, and the variance of the angular momentum $\mathrm{Var}(L_z)$, at the onset of the macrovortex decay.
    All the plots were obtained for $N=100$ bosons and $M=3$ orbitals after a trap quench $(p_1, p_2) = (-0.1, 0.05) \to (0.1, 0.0)$ of a 2-macrovortex with $g_0=0.2$ and $\Omega=0.65$.
    The dotted vertical line is a guide to the eye.
    }
    \label{fig:quadrupole}
\end{figure}

We first remark that each decay event is consistently preceded by the appearance of a nonzero quadrupole moment, as shown in Fig.~\ref{fig:quadrupole}, indicating that the breakup of the macrovortex is accompanied by a breaking of rotational symmetry.
In fact, as panel (b) shows, as soon as the macrovortex begins to break apart, angular momentum quantization is broken. 
Interestingly, the angular momentum first decreases before eventually monotonically increasing once the macrovortex has fully split into single vortices.
The macrovortex decay is also detected by the variance of the angular momentum, as panel (c) shows.

In retrospect, Fig.~\ref{fig:decay} already suggested a more subtle property of macrovortex stability.
For tighter traps (larger $p_1$), the macrovortex lifetime increases, roughly following a square-root or weak logarithmic law. 
Physically, this can be understood from the reduced phonon phase space: a tighter confinement increases the discrete energy spacing of radial modes~\cite{Dalfovo:1999}, limiting the number of available compressible excitations that can absorb angular momentum from the core. 
Conversely, for shallower traps, the macrovortices decay almost immediately after the quench\footnote{The rapid decay observed for very weak confinements ($p_1 \lesssim 0.1$) should, however, be interpreted with caution, as the condensate can expand toward the simulation boundaries and self-interfere due to the finite grid size.}.

\begin{figure}[t!]
    \centering
    \includegraphics[width=0.8\columnwidth]{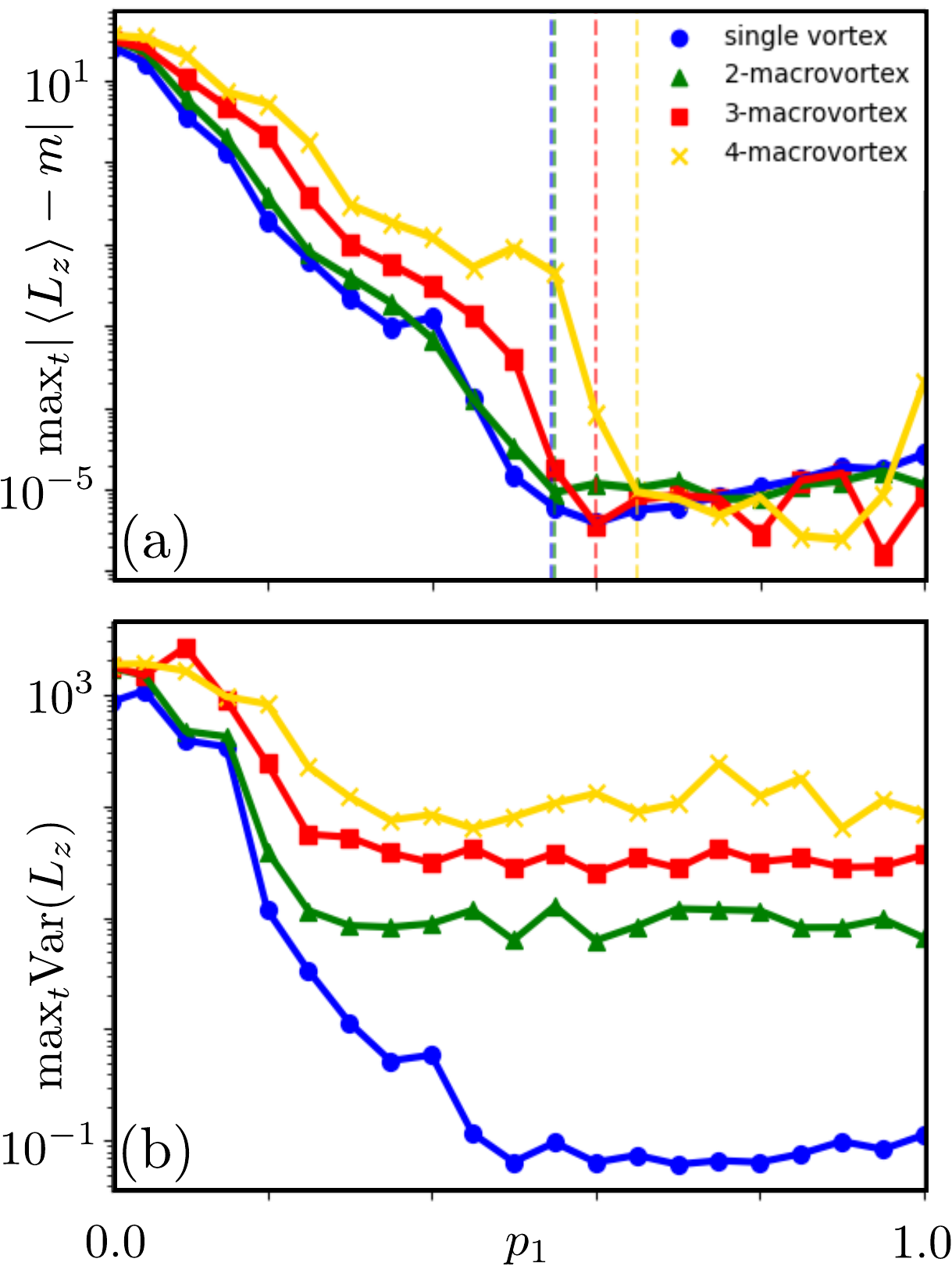}
    \caption{\textbf{Angular momentum as a function of trap parameter $p_1$.}
    (a) Maximal angular momentum deviation from vorticity $m$, plotted after a trap quench for different types of macrovortices, $m=1$ to $m=4$. The vertical dashed lines indicate the values of $p_1$ above which the deviation drops below $10^{-4}$.
    (b) Maximal angular momentum variance achieved after the same quench.
    The other parameters are as in Fig.~\ref{fig:decay}.
    }
    \label{fig:angular-momentum}
\end{figure}

The behavior of the angular momentum as a function of quenched trap parameter $p_1$, shown in Fig.~\ref{fig:angular-momentum}, confirms our intuition about the number of decay channels and macrovortex stability. 
Panel (a) shows the maximal deviation of the expectation value $\langle L_z \rangle$ from the quantized value $m$, while panel (b) presents the corresponding maximal variance $\mathrm{Var}(L_z)$. 
For sufficiently tight traps (large $p_1$), $\langle L_z \rangle$ remains nearly quantized, and fluctuations are suppressed.
This is in accordance with the rotational symmetry of the trap in the absence of a large number of decay channels.
As the trap becomes weaker, instead, both the deviation and the variance grow rapidly, signaling the onset of angular momentum sharing among the emerging vortices, likely mediated by phonons.

\begin{figure*}[t!]
    \centering
    \includegraphics[width=\textwidth]{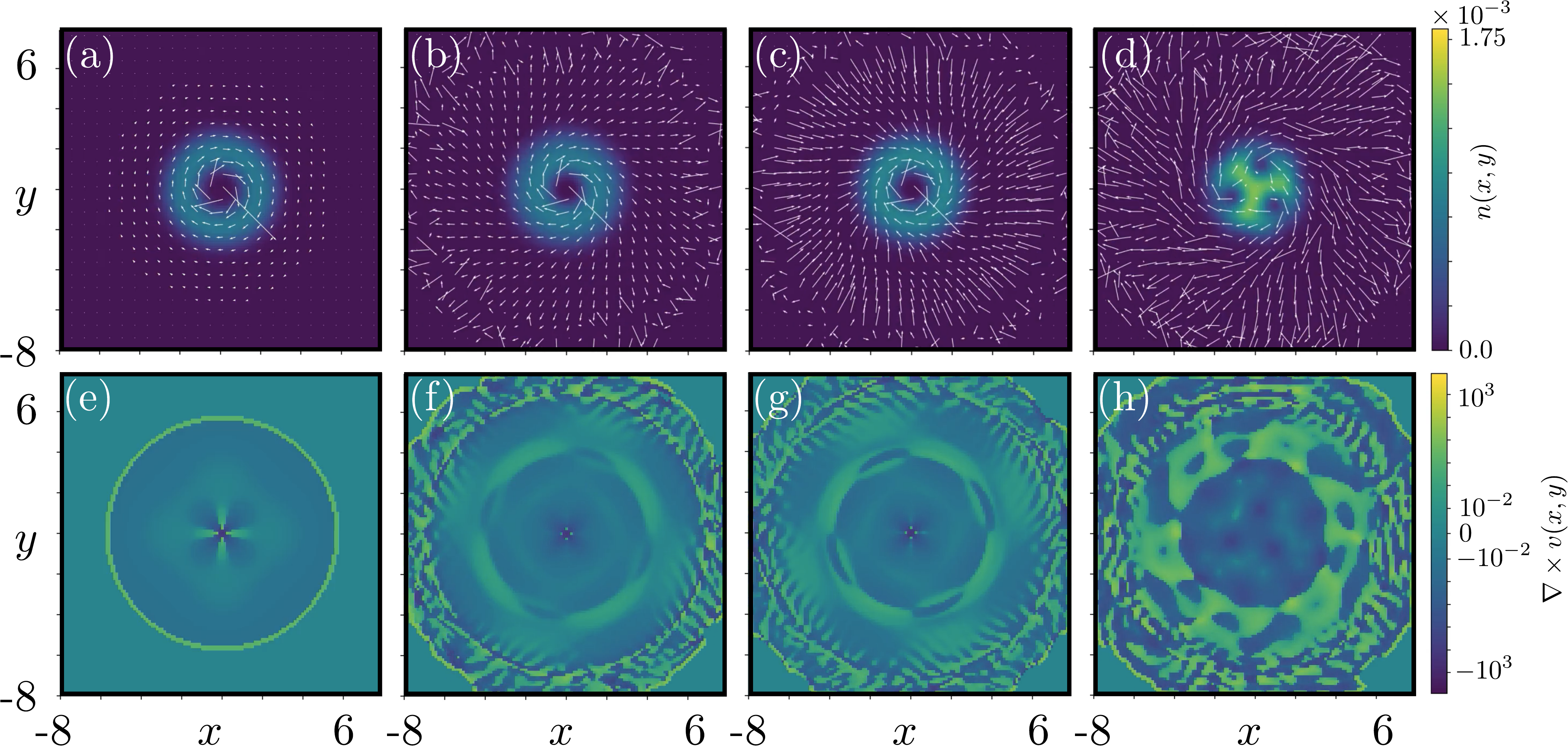}
    \caption{\textbf{Traces of vortex-phonon interactions.}
    Snapshot of the natural density $n(x,y)$ and velocity field $v(x,y)$ (top panels), and of the vorticity $\nabla \times v(x,y)$ for a trap quench $(p_1, p_2)=(-0.1, 0.05) \to (1.0, 0.0)$ with an initial 2-macrovortex obtained with $g_0=0.2$, $\Omega=0.65$.
    The panels show snapshots at times 
    (a),(e) $t=0$ (initial state), 
    (b),(f) $t=141$ (before decay),
    (c),(g) $t=142$ (before decay),
    (d),(h) $t=225$ (after decay).
    The results are obtained for $N=100$ bosons in $M=3$ orbitals.
    }
    \label{fig:phonon-ints}
\end{figure*}

To better visualize the role of phonons, Fig.~\ref{fig:phonon-ints} shows the density, velocity field, and corresponding vorticity for a representative trap quench starting from a 2-macrovortex.
The initial state show a purely solenoidal motion around the central core consistent with a macrovortex state.
After the quench, though, the condensate’s motion is not purely solenoidal anymore but exhibits alternating compressions and expansions, visible as faint breathing oscillations superimposed on the rotational flow. 
The velocity field (upper panels) also highlights how the flow intermittently redirects energy between the central region and the outer annulus, indicating the presence of compressible modes. 
This motion culminates in a separation of the macrovortex into multiple singly quantized vortices, each surrounded by a characteristic three-armed spiral flow pattern. 

The vorticity maps [bottom panels of Fig.~\ref{fig:phonon-ints}] provide a particularly intuitive picture of this process. 
They display concentric ring-like ripples that are continuously shed and reabsorbed by the central vortex core. These ripples correspond to propagating density modulations -- i.e. phonons -- that mediate energy transport within the condensate. 
The recurrent emission and reabsorption of such ripples signal that the macrovortex is constantly coupled to a fluctuating background of compressible excitations, which act as an effective dissipative bath even in the absence of explicit damping.

\begin{figure}[t!]
    \centering
    \includegraphics[width=0.9\columnwidth]{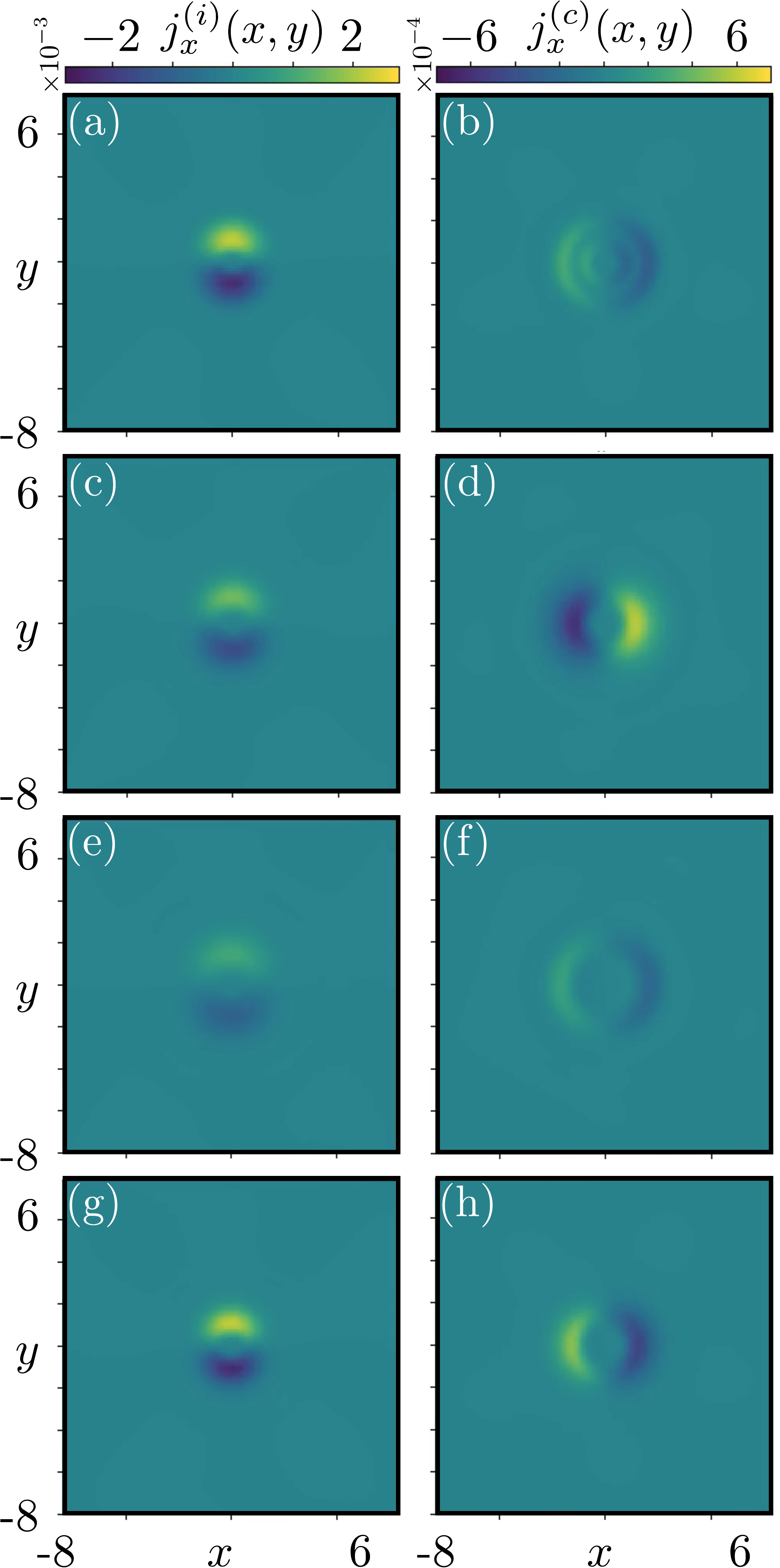}
    \caption{\textbf{Helmholtz current dynamics.}
    (a)-(h) Snapshots at four successive times of the $x$-components of the incompressible ($j^{(i)})(x,y)$, left column) and compressible ($j^{(c)})(x,y)$, right column) parts of the one-body current density after a Helmholtz decomposition.
    The panels show snapshots before the macrovortex decay, at times 
    (a)-(b) $t=127$,
    (c)-(d) $t=128$,
    (e)-(f) $t=129$,
    (g)-(h) $t=130$.
    All the plots are obtained after a trap quench $(p_1, p_2)=(-0.1, 0.05) \to (1.0, 0.0)$ with an initial 2-macrovortex prepared with $N=100$ bosons in $M=3$ orbitals for $g_0=0.2$, $\Omega=0.65$.
    }
    \label{fig:current-dyn}
\end{figure}

To quantify this interplay between vortices and phonons, we perform a Helmholtz decomposition of the one-body current density into incompressible (solenoidal) and compressible (phononic) components, $\mathbf{j} = \mathbf{j}^{(i)} + \mathbf{j}^{(c)}$. 
The $x$-components of these currents are displayed in Fig.~\ref{fig:current-dyn}(a–h) at four successive times before the decay (the $y$-components display an analogous behavior but for brevity they are not shown here).

The incompressible current $\mathbf{j}^{(i)}$ (left column) shows a nearly symmetric breathing motion with alternating regions of inflow and outflow, but without substantial deformation of the overall vortex pattern. 
In contrast, the compressible current $\mathbf{j}^{(c)}$ (right column) exhibits sign changes and concentric interference-like structures, reflecting the propagation of density waves throughout the condensate. 
These patterns represent sound waves emitted and reabsorbed by the macrovortex, providing direct evidence of the phonon-mediated nature of the instability. 
The appearance of ripple-like features and alternating flow directions in $\mathbf{j}^{(c)}$ confirms that a significant portion of the system’s kinetic energy is transiently stored in compressible modes before being transferred back to the vortical flow.

\begin{figure}[t!]
    \centering
    \includegraphics[width=0.9\columnwidth]{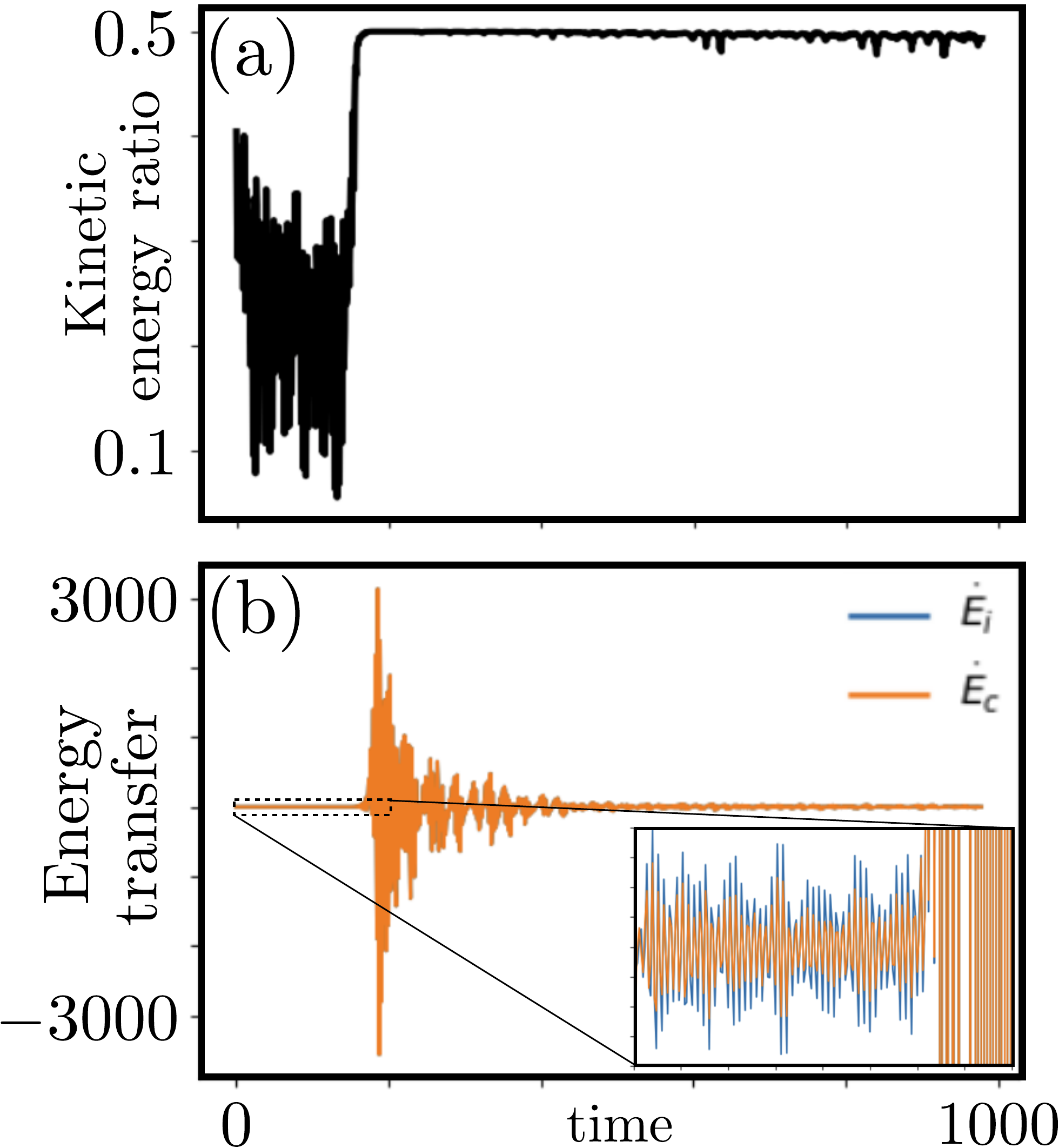}
    \caption{\textbf{Energy transfer after the trap quench.}
    (a) Dynamics of the ratio $R_c$ between incompressible and compressible kinetic energies obtained from the Helmholtz current fields.
    (b) Time derivatives of the energy components -- $\frac{\mathrm{d} E_i}{\mathrm{d} t}$, $\frac{\mathrm{d} E_c}{\mathrm{d} t}$ -- showing energy transfer between incompressible modes (vortices) and incompressible modes (phonons).
    All the plots are obtained after a trap quench $(p_1, p_2)=(-0.1, 0.05) \to (1.0, 0.0)$ with an initial 2-macrovortex prepared with $N=100$ bosons in $M=3$ orbitals for $g_0=0.2$, $\Omega=0.65$.
    }
    \label{fig:energy-dyn}
\end{figure}

This interpretation is further supported by the energy dynamics shown in Fig.~\ref{fig:energy-dyn}. 
The ratio $R_c = E_c / (E_i + E_c)$ between the compressible and incompressible kinetic energies [panel (a)] increases sharply just before the decay and stabilizes around $R_c \simeq 0.5$ after the breakup, indicating an almost equal sharing of energy between phonons and vortices. 
The time derivatives of the kinetic energy components [panel (b)] clearly demonstrate alternating energy transfer between $E_i$ and $E_c$, consistent with periodic exchange between rotational and vibrational motion. 
At the decay event, this exchange culminates in a sudden and pronounced transfer of energy from the vortical to the phononic sector, corresponding to the burst of sound waves emitted during vortex splitting. 
Afterward, the system settles into a quasi-stationary regime where both components fluctuate weakly around equilibrium.

\section{Conclusions and Outlook} 
\label{sec:conclusions}

We have presented a comprehensive study of the stability and dynamics of macrovortex states in a rotating Bose gas confined by a Mexican-hat potential with correlated multiorbital simulations. 

Our results reveal that quantum correlations, captured by increasing the number of orbitals $M$, qualitatively reshape the phase diagram and strongly influence the stability of macrovortex configurations. 
At the mean-field level, the transitions between superfluid, vortex-lattice, and macrovortex phases appear sharp, whereas including correlations smooths these boundaries and stabilizes coexistence regions where both types of vortices can coexist. 

The second key finding concerns the response of these states to quenches: while rotation or interaction quenches leave the macrovortices remarkably stable and merely excite clean, well-separated breathing modes, trap quenches inevitably lead to their decay. 

The third message is that macrovortex decay follows a universal, phonon-assisted mechanism in which energy and angular momentum are cyclically transferred from the rotational to the compressible sector. 
During this process, the macrovortex acts as an emitter of compressible excitations that progressively drain its rotational energy. 
As the phonon activity builds up, the coherence of the multiply charged core is disrupted, ultimately leading to its fragmentation into singly quantized vortices. 
After the decay, the system relaxes into a mixed steady state where vortical and phononic kinetic energies contribute in nearly equal proportions.

Looking ahead, our findings open several avenues for exploration. 
On the theoretical side, the clear frequency separation between macrovortices of different vorticity suggests a pathway toward information encoding in collective modes: each macrovortex could serve as a classical register, uniquely identified by its characteristic monopole frequency. 
Extending this concept into the quantum domain, one could envision coherent superpositions of counter-rotating states, such as $\ket{+m} + \ket{-m}$, thereby realizing a rotational qubit based on vorticity. 
Achieving this would require precise control over the trap geometry and interaction landscape to enable transitions between vortex states while suppressing uncontrolled phonon emission. 

On the other hand, our results from the trap quench indicate that macrovortex stability can be further tuned not only through external dissipation but also by adjusting the trap frequencies themselves, since these directly determine the number of accessible phononic channels and hence the available decay pathways.

From an experimental perspective, the signatures identified here -- such as the distinct breathing frequencies, the onset of quadrupole motion preceding decay, and the redistribution of energy between compressible and incompressible modes -- are all directly accessible with current imaging and spectroscopic techniques in ultracold atomic gases. 
Collective modes, including breathing and quadrupole oscillations, have been extensively measured using in situ absorption or phase-contrast imaging as well as trap-modulation spectroscopy~\cite{Chevy:2002, Bretin:2003}. 
The separation between compressible and incompressible kinetic-energy components can be experimentally characterized through Bragg spectroscopy~\cite{Stenger:1999, Steinhauer:2003}.

In particular, systems of rubidium atoms in ring traps or annular optical lattices provide a natural platform to test the predicted stability diagrams and decay dynamics.
Exploring these effects under controlled dissipative conditions, where the balance between phononic and vortical channels can be tuned externally, would offer a powerful means to probe and manipulate vortex–phonon coupling and energy transport in quantum fluids.

In summary, this work establishes macrovortices in rotating Mexican-hat potentials as a versatile platform for studying long-range coherence, collective excitations, and energy transfer in strongly driven superfluids. 
Beyond their fundamental significance, these states may offer new routes toward hybrid classical–quantum information encoding and the controlled manipulation of topological excitations in quantum gases.

\textit{Acknowledgements --}
We thank Leandro Alvares Machado for useful discussions.
This work was supported by the Swedish Research Council (2024-05213).
This work was supported by the São Paulo Research Foundation under Grants No. 2013/07276-1, No. 2024/04637-8, and No.2025/00547-7, and by the National Council for Scientific and Technological Development under Grant No. 386392/2024-2. Texas A\&M University is acknowledged.
Computation time at the High-Performance Computing Center Stuttgart (HLRS), on the Sunrise Compute Cluster of Stockholm University, and on the Euler cluster at the High-Performance Computing Center of ETH Zurich is gratefully acknowledged. 

\appendix
\section{MCTDH-X}
\label{app:MCTDHX}
In this appendix, we provide a summary of the multiconfigurational method used in this work.
All simulations were performed with the \textsc{MCTDH-X} software package, which implements the MultiConfigurational Time-Dependent Hartree method for indistinguishable particles (MCTDH-X)~\cite{Alon:2008, Lode:2012, Lode:2016, Fasshauer:2016, Lode:2020, Lin:2020, Molignini:2025-SciPost}.  
This approach provides a numerically exact solution of the time-dependent Schrödinger equation within a systematically improvable variational subspace.
The versatility of MCTDH-X is evident from its successful application across a wide variety of ultracold atomic systems.
These range from noninteracting systems~\cite{Xiang:2023}, to short-range interacting gases in various geometries and dynamical configurations~\cite{Beinke:2018, Roy:2018, Dutta:2019, Schaefer:2020, Lode:2021, Lode:2021-10, Debnath:2023, Roy:2023, Dutta:2023, Dutta:2024, Haldar:2024, Roy:2024-09, Roy:2024-11, Chatterjee:2024, Bhowmik:2025, Chakrabarti:2025-2}, to dipolar atoms and molecules~\cite{Fischer:2015, Chatterjee:2018, Chatterjee:2019, Bera:2019, Bera:2019-symm, Chatterjee:2020, Roy:2022, Hughes:2023, Bilinskaya:2024, Molignini:2024, Molignini:2024-2, Roy:2024-annals, Roy:2024-epjp, Molignini:2025-quasicryst1, Molignini:2025-quasicryst2, Chakrabarti:2025, Molignini:2025-JPCM}, and to ultracold gases coupled to optical cavities~\cite{Lode:2017, Lode:2018, Molignini:2018, Lin:2019, Lin2:2019, Lin:2020-PRA, Lin:2021, Molignini:2022, Rosa-Medina:2022, Ortuno-Gonzalez:2025}.
In this appendix, we focus exclusively on bosonic systems (MCTDHB), although MCTDH-X has also been extended to treat fermionic gases and spinor condensates.

The many-body wave function of $N$ bosons is expressed as a time-dependent linear combination of permanents,
\begin{equation}
|\Psi(t)\rangle = \sum_{\vec{n}} C_{\vec{n}}(t)\,|\vec{n};t\rangle,
\end{equation}
where each permanent $|\vec{n};t\rangle$ is built from $M$ time-dependent orbitals $\{\psi_k(\mathbf{r},t)\}$ and occupation numbers $\vec{n}=(n_1,\ldots,n_M)$ satisfying $\sum_k n_k=N$.  
Both the coefficients $\{C_{\vec{n}}(t)\}$ and the orbitals $\{\psi_k(\mathbf{r},t)\}$ evolve in time according to the Dirac–Frenkel variational principle, ensuring full self-consistency between the orbital basis and configuration space.

For $M=1$, this ansatz reduces to the Gross–Pitaevskii equation (mean-field limit). 
Increasing $M$ systematically incorporates many-body correlations and allows one to capture fragmentation, clustering, and other beyond-mean-field effects. 
In the limit $M\to\infty$, the expansion becomes exact.
In practice, convergence is achieved for finite $M$ once observables such as energy and density profiles cease to change upon increasing the number of orbitals. 
All calculations presented in this work with $N=100$ and $M=3$ are within the limit of what is computable with our numerical resources.

Ground states are obtained by imaginary-time propagation of the coupled equations. 
From the converged many-body state $|\Psi\rangle$, we compute observables presented in the main text, such as the one-body density
\begin{equation}
\rho(\mathbf{r}) = \langle \Psi | \hat{\Psi}^\dagger(\mathbf{r}) \hat{\Psi}(\mathbf{r}) | \Psi \rangle,
\end{equation}
and the spectral decomposition of the one-body reduced density matrix,
\begin{equation}
\rho^{(1)}(\mathbf{r},\mathbf{r}') = \sum_i \rho_i\,\phi_i^{(\mathrm{NO})}(\mathbf{r}) \phi_i^{(\mathrm{NO})*}(\mathbf{r}').
\end{equation}
Further methodological details and benchmarks of the MCTDH-X approach can be found in the literature.

\begin{figure}[t!]
    \centering
    \includegraphics[width=\columnwidth]{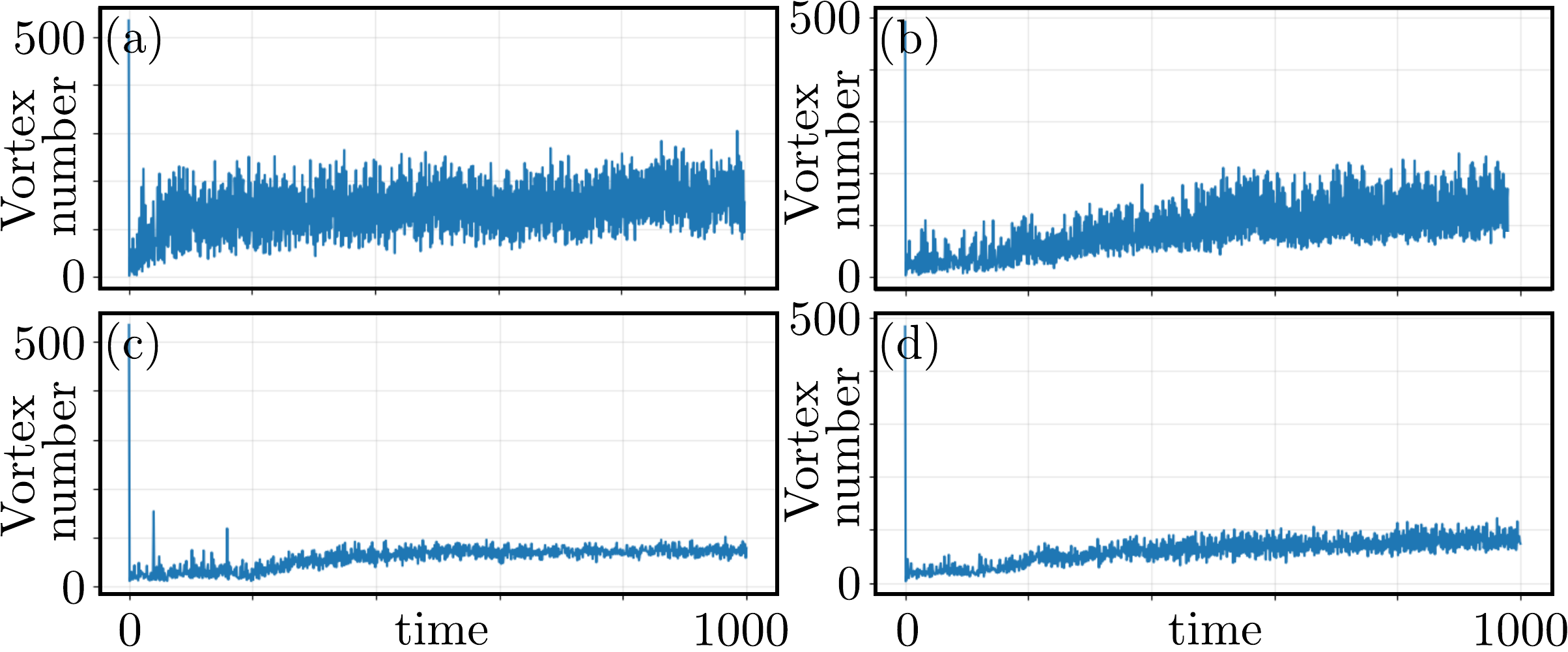}
    \caption{\textbf{Post-quench vortex generation.}
    Total number of vortices as a function of time in a system of $N=100$ bosons in $M=3$ orbitals, for a trap quench $(p_1, p_2)=(-0.1, 0.05) \to (1.0, 0.0)$.
    The different panels correspond to different initial states:
    (a) $g_0=0.5$, $\Omega=0.5$ (single-vortex),
    (b) $g_0=0.2$, $\Omega=0.65$ (2-macrovortex),
    (c) $g_0=0.2$, $\Omega=0.7$ (3-macrovortex),
    (d) $g_0=0.15$, $\Omega=0.75$ (4-macrovortex).
    }
    \label{fig:vortex-generation}
\end{figure}

\section{Units and system parameters}
\label{app:units}
Throughout our simulations, we employ dimensionless atomic units, setting $\hbar=m=1$. 
The two–dimensional isotropic trap used to control the system is
\begin{equation}
V(r)=p_1 r^2 + p_2 r^4 .
\end{equation}
If $p_2=0$ and $p_1>0$, the trap is purely harmonic,
\(
V(r)=\tfrac12\,\omega_{\rm h}^2 r^2
\)
with an effective frequency
\begin{equation}
\omega_{\rm h}=\sqrt{2\,p_1}.
\end{equation}
If $p_1<0$ and $p_2>0$, $V$ has a ring minimum at radius $r_0=\sqrt{\frac{-p_1}{2p_2}}$.
Expanding $V$ about $r_0$ and setting $r=r_0+\rho$ gives
\(
V(r)=V(r_0)+\tfrac12\,\omega_r^2\,\rho^2+\mathcal{O}(\rho^3)
\),
where the radial small–oscillation frequency follows from the curvature $\frac{d^2V}{dr^2}=2p_1+12p_2 r^2$ evaluated at $r_0$:
\begin{equation}
\omega_r=\sqrt{\,2p_1+12p_2 r_0^2\,}=\sqrt{-4p_1}=2\sqrt{|p_1|} .
\end{equation}
Note that while the \emph{radius} $r_0$ depends on both $(p_1,p_2)$, the \emph{radial stiffness} (and thus $\omega_r$) depends only on $p_1$ for this quartic form.

We define our dimensionless units using a chosen reference frequency $\omega_{\rm ref}$ obtained from the shape of the potential. 
For example, for the units of length, energy, time, and velocity, we find:
\begin{align}
L_0 &= \sqrt{\frac{\hbar}{m\omega_{\rm ref}}},\qquad
E_0 = \hbar\omega_{\rm ref}, \\
t_0 &= \frac{1}{\omega_{\rm ref}},\qquad
v_0 =\frac{L_0}{t_0}=\sqrt{\frac{\hbar\omega_{\rm ref}}{m}}.
\end{align}
Other units can be obtained analogously.

Note that in the trap quench protocols shown in the main text, $\omega_{\rm ref}$ is taken to be the \emph{post–quench} harmonic frequency $\omega_{\rm h}$; consequently, all times on the dynamical plots are already expressed in the post–quench unit $t_0=1/\omega_{\rm h}$.

In terms of parameter choices for our simulations, the
pre–quench Mexican–hat potential has parameters $(p_1,p_2)=(-0.1,0.05)$, hence $r_0=\sqrt{\frac{0.1}{0.1}}=1$ and $\omega_r=2\sqrt{0.1}\approx 0.6325$.
For the post-quench dynamics, $(p_1,p_2)=(0.1,0)$, and
$\omega_{\rm h}=\sqrt{0.2}\approx 0.4472$, $t_0=\frac{1}{\omega_{\rm h}}\approx 2.236$.

Unless otherwise stated, simulations were performed with $N=100$ bosons and $M=3$ orbitals. 
The grid extends over $x,y\in[-8L_0,8L_0]$ with $128$ grid points per spatial direction, corresponding to a resolution of $0.125\,L_0$.

\begin{figure}[t!]
    \centering
    \includegraphics[width=\columnwidth]{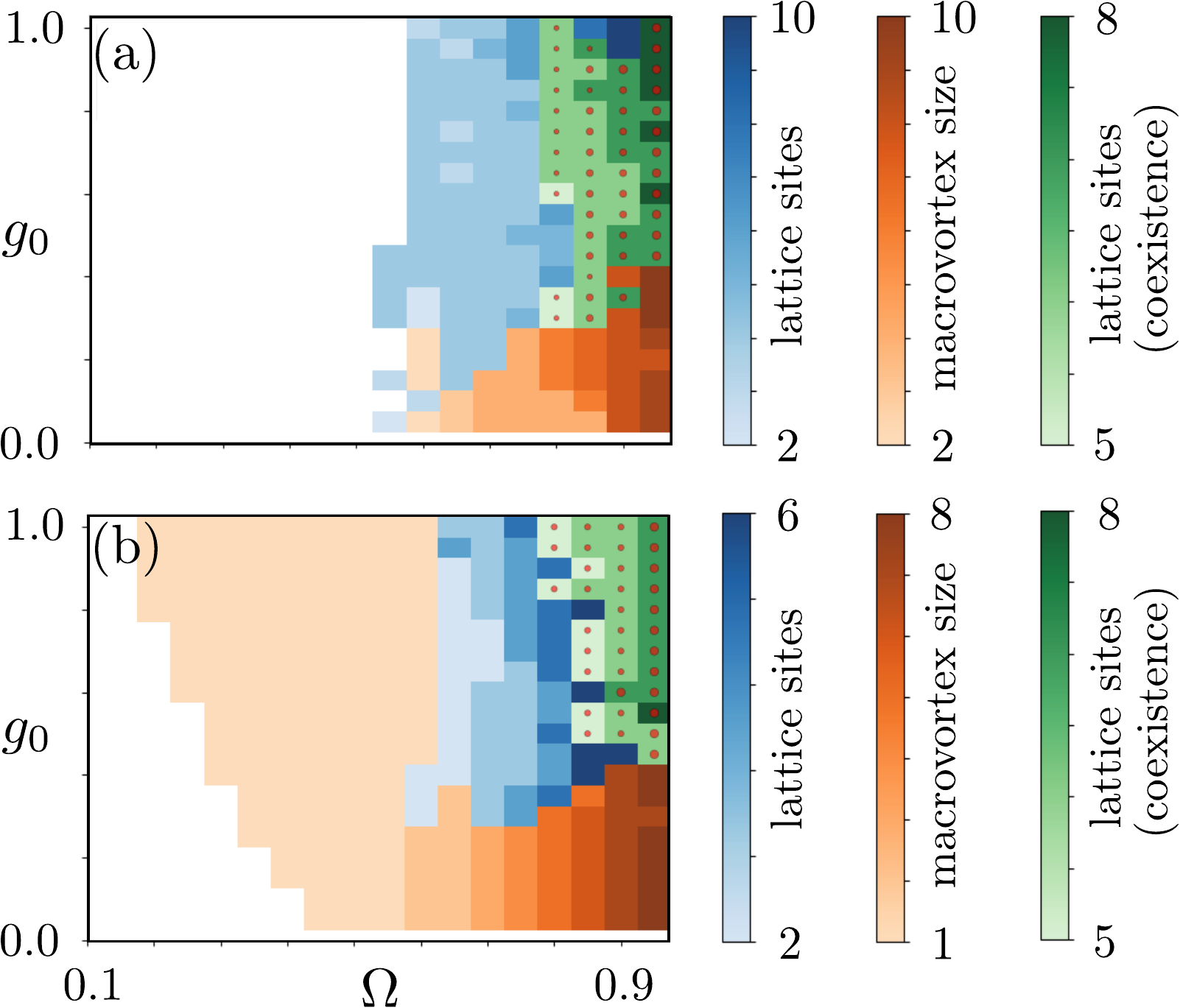}
    \caption{\textbf{Phase diagram for $N=100$ rotating bosons with narrow Gaussian interactions in a Mexican-hat potential.} The diagram is plotted as a function of angular velocity $\Omega$ and interaction strength $g_0$, plotted for (a) $M=1$ orbital (mean-field result), (b) $M=3$ orbitals. The different color schemes indicate the three possible phases: vortex lattice only (blue colorbar), central macrovortex only (orange colorbar), and coexistence of the two (green colorbar). For the region where the central macrovortex coexists with a surrounding vortex lattice, the red dots additionally mark the macrovortex size.}
    \label{fig:PD-Gaussian}
\end{figure}

\section{Vortex generation}
\label{app:vortex-generation}

Fig.~\ref{fig:vortex-generation} illustrates the typical post-quench dynamics following the transition from the Mexican-hat to the harmonic trap.  
The quench injects kinetic energy into the condensate, leading to a rapid proliferation of vortex–antivortex pairs.  
These are identified automatically from the phase winding of the most occupied orbital in the many-body wave function.  

A strong initial burst of vorticity occurs immediately after the quench, followed by a noisy but nearly stationary regime where the number of vortices fluctuates around a characteristic value.  
The magnitude of these fluctuations depends on the interaction strength $g_0$ and the rotation frequency $\Omega$.  
The behavior is consistent across different initial macrovortex configurations, confirming that the decay of the central macrovortex is not triggered by the generation of many vortex–antivortex pairs, and can instead related to phonon emission.

\section{Gaussian interactions}
\label{app:Gaussian}

\begin{figure}[t!]
    \centering
    \includegraphics[width=\columnwidth]{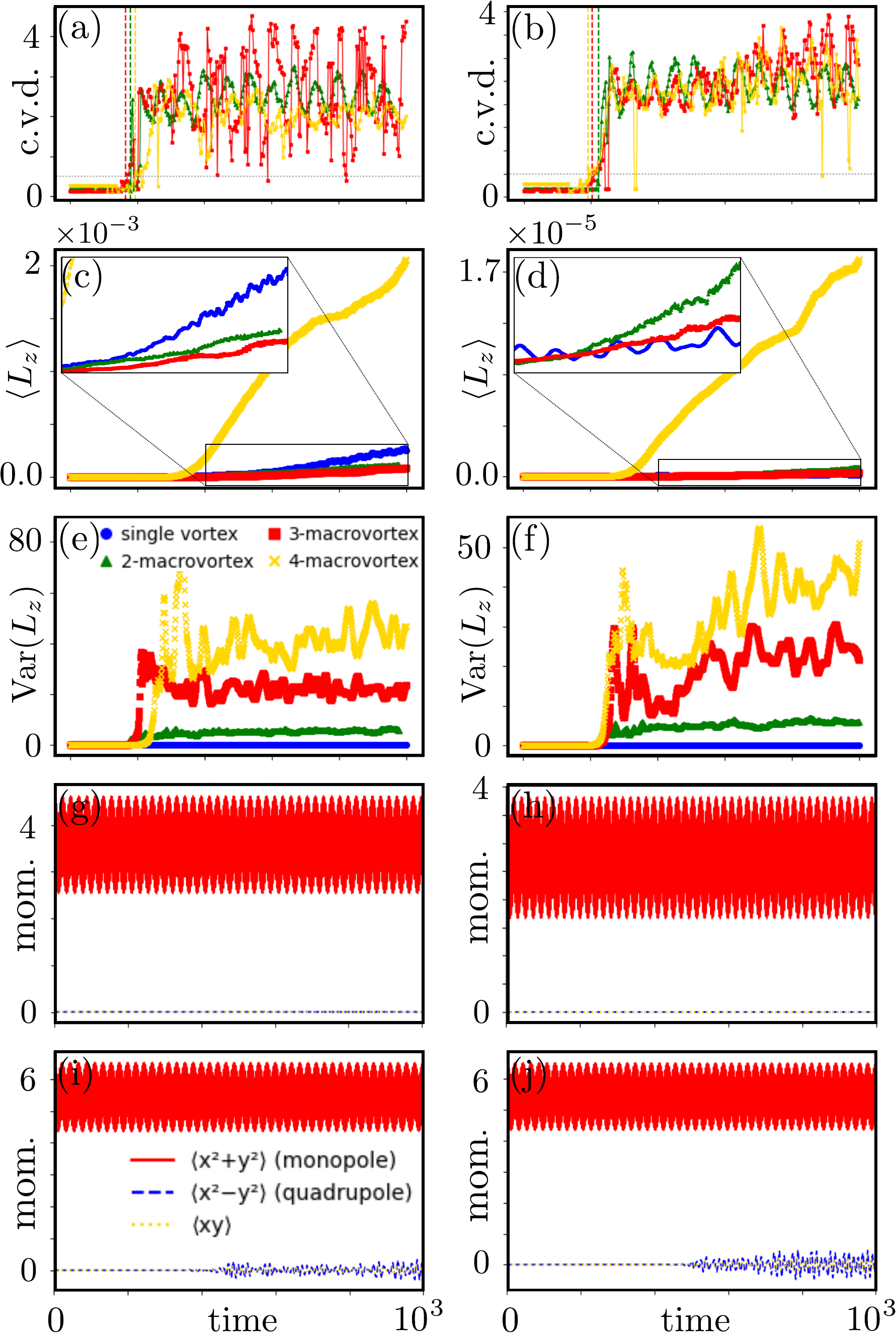}
    \caption{\textbf{Comparison of quench dynamics between contact and Gaussian interactions.}
    The left column depicts results for delta interactions $W(r,r')=g_0 \delta(r-r')$, while the right column shows results for narrow Gaussians $W(r,r')= \frac{g_0}{2\pi \sigma^2} \exp(-(r-r')^2/2 \sigma^2)$ with $\sigma=0.15$.
    (a)-(b): Central vortex distance (c.v.d.) as a function of time; common legend with panel (e).
    (c)-(d): Dynamics of expected angular momentum $\left< L_z \right>$; common legend with panel (e).
    (e)-(f): Dynamics of the angular momentum variance $\mathrm{Var}(L_z)$.
    (g)-(h): Dynamics of monopole and quadrupole moments for a single-vortex state; common legend with panel (i).
    (i)-(j): Dynamics of monopole and quadrupole moments for a 4-macrovortex state.
    All initial states are prepared with $N=100$ particles and $M=3$ orbitals in the same Mexican-hat potential with $p_1=-0.1$, $p_2=0.05$.  
    }
    \label{fig:comparison-Gaussian}
\end{figure}
In two dimensions, the use of contact ($\delta$-function) interactions is known to present a mathematical subtlety.  
A bare $\delta$ potential in 2D is not self-adjoint and leads to ultraviolet divergences when inserted into the many-body Hamiltonian, requiring renormalization to correctly reproduce the physical $s$-wave scattering amplitude.  
Equivalently, the 2D scattering length depends logarithmically on the collision energy, so that the naive replacement $W(\mathbf{r},\mathbf{r}') = g_0\,\delta(\mathbf{r}-\mathbf{r}')$ does not correspond to a well-defined interaction strength in the continuum limit.  
In practice, this pathology affects only the very short-distance (ultraviolet) behavior, while the long-wavelength (infrared) physics remains accurately captured by a contact potential provided the relevant length scales -- such as the healing length or vortex-core size -- are much larger than the effective interaction range.  

To verify the validity of using delta-contact interactions in two-dimensional simulations, we performed an additional set of calculations with narrow Gaussian interactions,
\begin{equation}
W(\mathbf{r},\mathbf{r}') = 
\frac{g_0}{2\pi \sigma^2}
\exp\!\left[-\frac{|\mathbf{r}-\mathbf{r}'|^2}{2\sigma^2}\right],
\end{equation}
with width $\sigma = 0.15$. 

The results we present in this appendix pertain to both the ground-state physics and the post-quench dynamics.
The phase diagrams obtained with Gaussian interactions are shown in Fig.~\ref{fig:PD-Gaussian}.
They demonstrate that the equilibrium phase diagrams for the initial states are qualitatively identical to those obtained for contact interactions $W(\mathbf{r},\mathbf{r}') = g_0 \delta(\mathbf{r}-\mathbf{r}')$.
The same three phases -- vortex lattice (blue), central macrovortex (orange), and coexistence (green with red dots) -- emerge with almost unchanged phase boundaries.
In particular, we observe the same key features when moving from $M=1$ to $M=3$ simulations: (i) existence of single-vortex states at lower values of $\Omega$ for $M=3$, (ii) shift towards higher value of $g_0$ for the vortex lattice phase when $M=3$, and (iii) smoother transitions when correlations are included with $M=3$.

Similarly, the comparison in Fig.~\ref{fig:comparison-Gaussian} for dynamical quantities post-quench reveals that the angular momentum, its variance, and collective mode dynamics behave analogously in both cases.  
This confirms that our 2D $\delta$-interaction model faithfully reproduces the low-energy physics of systems with short-range Gaussian interactions.

\bibliography{biblio}

@article{Cooper_2019,
	author = {Cooper, N. R. and Dalibard, J. and Spielman, I. B.},
	doi = {10.1103/revmodphys.91.015005},
	issn = {1539-0756},
	journal = {Reviews of Modern Physics},
	month = mar,
	number = {1},
	publisher = {American Physical Society (APS)},
	title = {Topological bands for ultracold atoms},
	url = {http://dx.doi.org/10.1103/RevModPhys.91.015005},
	volume = {91},
	year = {2019}}

@article{RevModPhys.71.463,
	author = {Dalfovo, Franco and Giorgini, Stefano and Pitaevskii, Lev P. and Stringari, Sandro},
	doi = {10.1103/RevModPhys.71.463},
	issue = {3},
	journal = {Rev. Mod. Phys.},
	month = {Apr},
	numpages = {0},
	pages = {463--512},
	publisher = {American Physical Society},
	title = {Theory of Bose-Einstein condensation in trapped gases},
	url = {https://link.aps.org/doi/10.1103/RevModPhys.71.463},
	volume = {71},
	year = {1999}}

@article{RevModPhys.79.235,
	author = {Fort\'agh, J\'ozsef and Zimmermann, Claus},
	doi = {10.1103/RevModPhys.79.235},
	issue = {1},
	journal = {Rev. Mod. Phys.},
	month = {Feb},
	numpages = {0},
	pages = {235--289},
	publisher = {American Physical Society},
	title = {Magnetic microtraps for ultracold atoms},
	url = {https://link.aps.org/doi/10.1103/RevModPhys.79.235},
	volume = {79},
	year = {2007}}

@article{RevModPhys.80.1215,
	author = {Giorgini, Stefano and Pitaevskii, Lev P. and Stringari, Sandro},
	doi = {10.1103/RevModPhys.80.1215},
	issue = {4},
	journal = {Rev. Mod. Phys.},
	month = {Oct},
	numpages = {0},
	pages = {1215--1274},
	publisher = {American Physical Society},
	title = {Theory of ultracold atomic Fermi gases},
	url = {https://link.aps.org/doi/10.1103/RevModPhys.80.1215},
	volume = {80},
	year = {2008}}

@article{Pelucchi,
	author = {E. Pelucchi and G. Fagas and I. Aharonovich and D. Englund and E. Figueroa and Q. Gong and H. Hannes and J. Liu and C. Y. Lu and N. Matsuda and et al.},
	doi = {10.1038/s42254-021-00398-z},
	journal = {Nature Reviews Physics},
	pages = {194--208},
	title = {The potential and global outlook of integrated photonics for quantum technologies},
	url = {https://doi.org/10.1038/s42254-021-00398-z},
	volume = {4},
	year = {2022}}

@article{Debnath,
	author = {Debnath, S. and Linke, N. and Figgatt, C. et al.},
	doi = {10.1038/nature18648},
	journal = {Nature},
	pages = {63--66},
	title = {Demonstration of a small programmable quantum computer with atomic qubits},
	url = {https://doi.org/10.1038/nature18648},
	volume = {536},
	year = {2016}}

@article{Bloch:2012,
	author = {Bloch, I. and Dalibard, J. and Nascimb{\`e}ne, S.},
	doi = {10.1038/nphys2259},
	journal = {Nature Phys},
	pages = {267--276},
	title = {Quantum simulations with ultracold quantum gases},
	url = {https://doi.org/10.1038/nphys2259},
	volume = {8},
	year = {2012}}

@article{Yago_Malo_2024,
	author = {Yago Malo, Jorge and Lepori, Luca and Gentini, Laura and Chiofalo, Maria Luisa (Maril{\`u})},
	doi = {10.3390/technologies12050064},
	issn = {2227-7080},
	journal = {Technologies},
	month = may,
	number = {5},
	pages = {64},
	publisher = {MDPI AG},
	title = {Atomic Quantum Technologies for Quantum Matter and Fundamental Physics Applications},
	url = {http://dx.doi.org/10.3390/technologies12050064},
	volume = {12},
	year = {2024}}

@article{Anderson,
	author = {Anderson MH and Ensher JR and Matthews MR and Wieman CE and Cornell EA.},
	doi = {10.1126/science.269.5221.198},
	journal = {Science},
	pages = {198--201},
	title = {Observation of bose-einstein condensation in a dilute atomic vapor},
	url = {https://doi.org/10.1126/science.269.5221.198},
	volume = {269},
	year = {1995}}

@article{Gaunt_2013,
	author = {Gaunt, Alexander L. and Schmidutz, Tobias F. and Gotlibovych, Igor and Smith, Robert P. and Hadzibabic, Zoran},
	doi = {10.1103/physrevlett.110.200406},
	issn = {1079-7114},
	journal = {Physical Review Letters},
	month = may,
	number = {20},
	publisher = {American Physical Society (APS)},
	title = {Bose-Einstein Condensation of Atoms in a Uniform Potential},
	url = {http://dx.doi.org/10.1103/PhysRevLett.110.200406},
	volume = {110},
	year = {2013}}

@article{Greiner2002QuantumPT,
	author = {Markus Greiner and Olaf Mandel and Tilman Esslinger and Theodor W. H{\"a}nsch and Immanuel Bloch},
	journal = {Nature},
	pages = {39-44},
	title = {Quantum phase transition from a superfluid to a Mott insulator in a gas of ultracold atoms},
	url = {https://api.semanticscholar.org/CorpusID:4411344},
	volume = {415},
	year = {2002}}

@article{Bloch:2005,
	author = {Bloch, I.},
	doi = {10.1038/nphys138},
	journal = {Nature Phys},
	pages = {23-30},
	title = {Ultracold quantum gases in optical lattices.},
	url = {https://doi.org/10.1038/nphys138},
	volume = {1},
	year = {2005}}

@article{PhysRevA.73.013603,
	author = {Cozzini, M. and Jackson, B. and Stringari, S.},
	doi = {10.1103/PhysRevA.73.013603},
	issue = {1},
	journal = {Phys. Rev. A},
	month = {Jan},
	numpages = {10},
	pages = {013603},
	publisher = {American Physical Society},
	title = {Vortex signatures in annular Bose-Einstein condensates},
	url = {https://link.aps.org/doi/10.1103/PhysRevA.73.013603},
	volume = {73},
	year = {2006}}

@article{PhysRevA.59.2990,
	author = {Salasnich, L. and Parola, A. and Reatto, L.},
	doi = {10.1103/PhysRevA.59.2990},
	issue = {4},
	journal = {Phys. Rev. A},
	month = {Apr},
	numpages = {0},
	pages = {2990--2995},
	publisher = {American Physical Society},
	title = {Bosons in a toroidal trap: Ground state and vortices},
	url = {https://link.aps.org/doi/10.1103/PhysRevA.59.2990},
	volume = {59},
	year = {1999}}

@article{PhysRevA.69.033608,
	author = {Aftalion, Amandine and Danaila, Ionut},
	doi = {10.1103/PhysRevA.69.033608},
	issue = {3},
	journal = {Phys. Rev. A},
	month = {Mar},
	numpages = {6},
	pages = {033608},
	publisher = {American Physical Society},
	title = {Giant vortices in combined harmonic and quartic traps},
	url = {https://link.aps.org/doi/10.1103/PhysRevA.69.033608},
	volume = {69},
	year = {2004}}

@article{Ramanathan_2011,
	author = {Ramanathan, A. and Wright, K. C. and Muniz, S. R. and Zelan, M. and Hill, W. T. and Lobb, C. J. and Helmerson, K. and Phillips, W. D. and Campbell, G. K.},
	doi = {10.1103/physrevlett.106.130401},
	issn = {1079-7114},
	journal = {Physical Review Letters},
	month = mar,
	number = {13},
	publisher = {American Physical Society (APS)},
	title = {Superflow in a Toroidal Bose-Einstein Condensate: An Atom Circuit with a Tunable Weak Link},
	url = {http://dx.doi.org/10.1103/PhysRevLett.106.130401},
	volume = {106},
	year = {2011}}

@article{Corman_2014,
	author = {Corman, L. and Chomaz, L. and Bienaim{\'e}, T. and Desbuquois, R. and Weitenberg, C. and Nascimb{\`e}ne, S. and Dalibard, J. and Beugnon, J.},
	doi = {10.1103/physrevlett.113.135302},
	issn = {1079-7114},
	journal = {Physical Review Letters},
	month = sep,
	number = {13},
	publisher = {American Physical Society (APS)},
	title = {Quench-Induced Supercurrents in an Annular Bose Gas},
	url = {http://dx.doi.org/10.1103/PhysRevLett.113.135302},
	volume = {113},
	year = {2014}}

@article{Navez_2016,
	author = {Navez, P and Pandey, S and Mas, H and Poulios, K and Fernholz, T and Klitzing, W von},
	doi = {10.1088/1367-2630/18/7/075014},
	journal = {New Journal of Physics},
	month = {jul},
	number = {7},
	pages = {075014},
	publisher = {IOP Publishing},
	title = {Matter-wave interferometers using TAAP rings},
	url = {https://doi.org/10.1088/1367-2630/18/7/075014},
	volume = {18},
	year = {2016}}

@article{PhysRevLett.95.010402,
	author = {Albiez, Michael and Gati, Rudolf and F\"olling, Jonas and Hunsmann, Stefan and Cristiani, Matteo and Oberthaler, Markus K.},
	doi = {10.1103/PhysRevLett.95.010402},
	issue = {1},
	journal = {Phys. Rev. Lett.},
	month = {Jun},
	numpages = {4},
	pages = {010402},
	publisher = {American Physical Society},
	title = {Direct Observation of Tunneling and Nonlinear Self-Trapping in a Single Bosonic Josephson Junction},
	url = {https://link.aps.org/doi/10.1103/PhysRevLett.95.010402},
	volume = {95},
	year = {2005}}

@article{Kwon_2020,
	author = {Kwon, W. J. and Del Pace, G. and Panza, R. and Inguscio, M. and Zwerger, W. and Zaccanti, M. and Scazza, F. and Roati, G.},
	doi = {10.1126/science.aaz2463},
	issn = {1095-9203},
	journal = {Science},
	month = jul,
	number = {6499},
	pages = {84--88},
	publisher = {American Association for the Advancement of Science (AAAS)},
	title = {Strongly correlated superfluid order parameters from dc Josephson supercurrents},
	url = {http://dx.doi.org/10.1126/science.aaz2463},
	volume = {369},
	year = {2020}}

@article{PhysRevLett.111.205301,
	author = {Ryu, C. and Blackburn, P. W. and Blinova, A. A. and Boshier, M. G.},
	doi = {10.1103/PhysRevLett.111.205301},
	issue = {20},
	journal = {Phys. Rev. Lett.},
	month = {Nov},
	numpages = {5},
	pages = {205301},
	publisher = {American Physical Society},
	title = {Experimental Realization of Josephson Junctions for an Atom SQUID},
	url = {https://link.aps.org/doi/10.1103/PhysRevLett.111.205301},
	volume = {111},
	year = {2013}}

@article{PhysRevLett.99.260401,
	author = {Ryu, C. and Andersen, M. F. and Clad\'e, P. and Natarajan, Vasant and Helmerson, K. and Phillips, W. D.},
	doi = {10.1103/PhysRevLett.99.260401},
	issue = {26},
	journal = {Phys. Rev. Lett.},
	month = {Dec},
	numpages = {4},
	pages = {260401},
	publisher = {American Physical Society},
	title = {Observation of Persistent Flow of a Bose-Einstein Condensate in a Toroidal Trap},
	url = {https://link.aps.org/doi/10.1103/PhysRevLett.99.260401},
	volume = {99},
	year = {2007}}

@article{PhysRevA.86.013629,
	author = {Moulder, Stuart and Beattie, Scott and Smith, Robert P. and Tammuz, Naaman and Hadzibabic, Zoran},
	doi = {10.1103/PhysRevA.86.013629},
	issue = {1},
	journal = {Phys. Rev. A},
	month = {Jul},
	numpages = {7},
	pages = {013629},
	publisher = {American Physical Society},
	title = {Quantized supercurrent decay in an annular Bose-Einstein condensate},
	url = {https://link.aps.org/doi/10.1103/PhysRevA.86.013629},
	volume = {86},
	year = {2012}}

@article{RevModPhys.80.885,
	author = {Bloch, Immanuel and Dalibard, Jean and Zwerger, Wilhelm},
	doi = {10.1103/RevModPhys.80.885},
	issue = {3},
	journal = {Rev. Mod. Phys.},
	month = {Jul},
	numpages = {0},
	pages = {885--964},
	publisher = {American Physical Society},
	title = {Many-body physics with ultracold gases},
	url = {https://link.aps.org/doi/10.1103/RevModPhys.80.885},
	volume = {80},
	year = {2008}}

@article{Carollo_2022,
	author = {Carollo, R. A. and Aveline, D. C. and Rhyno, B. and Vishveshwara, S. and Lannert, C. and Murphree, J. D. and Elliott, E. R. and Williams, J. R. and Thompson, R. J. and Lundblad, N.},
	doi = {10.1038/s41586-022-04639-8},
	issn = {1476-4687},
	journal = {Nature},
	month = may,
	number = {7913},
	pages = {281--286},
	publisher = {Springer Science and Business Media LLC},
	title = {Observation of ultracold atomic bubbles in orbital microgravity},
	url = {http://dx.doi.org/10.1038/s41586-022-04639-8},
	volume = {606},
	year = {2022}}

@article{Aveline,
	author = {Aveline, D.C. and Williams and J.R. and Elliott and E.R. et al.},
	doi = {10.1038/s41586-020-2346-1},
	journal = {Nature},
	pages = {193--197},
	title = {Observation of Bose--Einstein condensates in an Earth-orbiting research lab},
	url = {https://doi.org/10.1038/s41586-020-2346-1},
	volume = {582},
	year = {2020}}

@article{Lundblad_2023,
	author = {Lundblad, Nathan and Aveline, David C and Bala{\v z}, Antun and Bentine, Elliot and Bigelow, Nicholas P and Boegel, Patrick and Efremov, Maxim A and Gaaloul, Naceur and Meister, Matthias and Olshanii, Maxim and S{\'a} de Melo, Carlos A R and Tononi, Andrea and Vishveshwara, Smitha and White, Angela C and Wolf, Alexander and Garraway, Barry M},
	doi = {10.1088/2058-9565/acb1cf},
	issn = {2058-9565},
	journal = {Quantum Science and Technology},
	month = feb,
	number = {2},
	pages = {024003},
	publisher = {IOP Publishing},
	title = {Perspective on quantum bubbles in microgravity},
	url = {http://dx.doi.org/10.1088/2058-9565/acb1cf},
	volume = {8},
	year = {2023}}

@article{PhysRevLett.86.1195,
	author = {Zobay, O. and Garraway, B. M.},
	doi = {10.1103/PhysRevLett.86.1195},
	issue = {7},
	journal = {Phys. Rev. Lett.},
	month = {Feb},
	numpages = {0},
	pages = {1195--1198},
	publisher = {American Physical Society},
	title = {Two-Dimensional Atom Trapping in Field-Induced Adiabatic Potentials},
	url = {https://link.aps.org/doi/10.1103/PhysRevLett.86.1195},
	volume = {86},
	year = {2001}}

@article{PhysRevA.69.023605,
	author = {Zobay, O. and Garraway, B. M.},
	doi = {10.1103/PhysRevA.69.023605},
	issue = {2},
	journal = {Phys. Rev. A},
	month = {Feb},
	numpages = {15},
	pages = {023605},
	publisher = {American Physical Society},
	title = {Atom trapping and two-dimensional Bose-Einstein condensates in field-induced adiabatic potentials},
	url = {https://link.aps.org/doi/10.1103/PhysRevA.69.023605},
	volume = {69},
	year = {2004}}

@article{Arazo_2021,
	author = {Arazo, Maria and Mayol, Ricardo and Guilleumas, Montserrat},
	doi = {10.1088/1367-2630/ac37c9},
	issn = {1367-2630},
	journal = {New Journal of Physics},
	month = nov,
	number = {11},
	pages = {113040},
	publisher = {IOP Publishing},
	title = {Shell-shaped condensates with gravitational sag: contact and dipolar interactions},
	url = {http://dx.doi.org/10.1088/1367-2630/ac37c9},
	volume = {23},
	year = {2021}}

@article{Garraway_2016,
	author = {Garraway, Barry M and Perrin, H{\'e}l{\`e}ne},
	doi = {10.1088/0953-4075/49/17/172001},
	journal = {Journal of Physics B: Atomic, Molecular and Optical Physics},
	month = {aug},
	number = {17},
	pages = {172001},
	publisher = {IOP Publishing},
	title = {Recent developments in trapping and manipulation of atoms with adiabatic potentials},
	url = {https://doi.org/10.1088/0953-4075/49/17/172001},
	volume = {49},
	year = {2016}}

@article{Colombe,
	author = {Y. Colombe and B. Mercier and H. Perrin and V. Lorent},
	doi = {10.1051/jp4:2004116013},
	journal = {J. Phys. IV France EDP Science},
	title = {Loading a dressed Zeeman trap with cold atoms},
	url = {https://doi.org/10.1051/jp4:2004116013},
	volume = {116},
	year = {2003}}

@article{Morizot_2007,
	author = {Morizot, O and Garrido Alzar, C L and Pottie, P-E and Lorent, V and Perrin, H},
	doi = {10.1088/0953-4075/40/20/004},
	journal = {Journal of Physics B: Atomic, Molecular and Optical Physics},
	month = {oct},
	number = {20},
	pages = {4013},
	title = {Trapping and cooling of rf-dressed atoms in a quadrupole magnetic field},
	url = {https://doi.org/10.1088/0953-4075/40/20/004},
	volume = {40},
	year = {2007}}

@article{PhysRevA.74.023616,
	author = {White, M. and Gao, H. and Pasienski, M. and DeMarco, B.},
	doi = {10.1103/PhysRevA.74.023616},
	issue = {2},
	journal = {Phys. Rev. A},
	month = {Aug},
	numpages = {4},
	pages = {023616},
	publisher = {American Physical Society},
	title = {Bose-Einstein condensates in rf-dressed adiabatic potentials},
	url = {https://link.aps.org/doi/10.1103/PhysRevA.74.023616},
	volume = {74},
	year = {2006}}

@article{PhysRevA.81.031402,
	author = {Gildemeister, M. and Nugent, E. and Sherlock, B. E. and Kubasik, M. and Sheard, B. T. and Foot, C. J.},
	doi = {10.1103/PhysRevA.81.031402},
	issue = {3},
	journal = {Phys. Rev. A},
	month = {Mar},
	numpages = {4},
	pages = {031402},
	publisher = {American Physical Society},
	title = {Trapping ultracold atoms in a time-averaged adiabatic potential},
	url = {https://link.aps.org/doi/10.1103/PhysRevA.81.031402},
	volume = {81},
	year = {2010}}

@article{PhysRevA.85.053401,
	author = {Gildemeister, M. and Sherlock, B. E. and Foot, C. J.},
	doi = {10.1103/PhysRevA.85.053401},
	issue = {5},
	journal = {Phys. Rev. A},
	month = {May},
	numpages = {6},
	pages = {053401},
	publisher = {American Physical Society},
	title = {Techniques to cool and rotate Bose-Einstein condensates in time-averaged adiabatic potentials},
	url = {https://link.aps.org/doi/10.1103/PhysRevA.85.053401},
	volume = {85},
	year = {2012}}

@article{Ryu,
	author = {Ryu, C. and Samson, E.C. and Boshier, M.G},
	doi = {10.1038/s41467-020-17185-6},
	journal = {Nat Commun},
	pages = {3338},
	title = {Quantum interference of currents in an atomtronic SQUID},
	url = {https://doi.org/10.1038/s41467-020-17185-6},
	volume = {11},
	year = {2020}}

@article{PhysRevLett.110.025302,
	author = {Wright, K. C. and Blakestad, R. B. and Lobb, C. J. and Phillips, W. D. and Campbell, G. K.},
	doi = {10.1103/PhysRevLett.110.025302},
	issue = {2},
	journal = {Phys. Rev. Lett.},
	month = {Jan},
	numpages = {5},
	pages = {025302},
	publisher = {American Physical Society},
	title = {Driving Phase Slips in a Superfluid Atom Circuit with a Rotating Weak Link},
	url = {https://link.aps.org/doi/10.1103/PhysRevLett.110.025302},
	volume = {110},
	year = {2013}}

@article{Herve_2021,
	author = {Mathieu de Goër de Herve and Yanliang Guo and Camilla De Rossi and Avinash Kumar and Thomas Badr and Romain Dubessy and Laurent Longchambon and Hélène Perrin},
	journal = {Journal of Physics B: Atomic, Molecular and Optical Physics},
	doi = {10.1088/1361-6455/ac0579},
	month = {jun},
	number = {12},
	pages = {125302},
	publisher = {IOP Publishing},
	url = {https://doi.org/10.1088/1361-6455/ac0579},
	volume = {54},
	year = {2021}}

@article{Tomishiyo_2024,
	author = {Tomishiyo, Guilherme and Madeira, Lucas and Caracanhas, M{\^o}nica A.},
	doi = {10.1063/5.0214294},
	issn = {1089-7666},
	journal = {Physics of Fluids},
	month = jun,
	number = {6},
	publisher = {AIP Publishing},
	title = {Superfluid excitations in rotating two-dimensional ring traps},
	url = {http://dx.doi.org/10.1063/5.0214294},
	volume = {36},
	year = {2024}}

@article{Brito_2020,
	author = {Brito, Leonardo and Andriati, Alex and Tomio, Lauro and Gammal, Arnaldo},
	doi = {10.1103/physreva.102.063330},
	issn = {2469-9934},
	journal = {Physical Review A},
	month = dec,
	number = {6},
	publisher = {American Physical Society (APS)},
	title = {Breakup of rotating asymmetric quartic-quadratic trapped condensates},
	url = {http://dx.doi.org/10.1103/PhysRevA.102.063330},
	volume = {102},
	year = {2020}}

@article{Adhikari_2019,
	author = {Adhikari, S K},
	doi = {10.1088/1612-202x/ab2d2c},
	issn = {1612-202X},
	journal = {Laser Physics Letters},
	month = jul,
	number = {8},
	pages = {085501},
	publisher = {IOP Publishing},
	title = {Stable controllable giant vortex in a trapped Bose--Einstein condensate},
	url = {http://dx.doi.org/10.1088/1612-202X/ab2d2c},
	volume = {16},
	year = {2019}}

@article{Chatterjee:2024,
	author = {B. Chatterjee},
	doi = {10.1007/s12043-023-02710-1},
	journal = {Pramana -- J. Phys.},
	pages = {34},
	title = {Vortex states in rotating Bose--Einstein condensates beyond the mean-field regime},
	url = {https://doi.org/10.1007/s12043-023-02710-1},
	volume = {98},
	year = {2024}}

@article{Streltsov:2006,
	author = {Alexej I. Streltsov and Ofir E. Alon and Lorenz S. Cederbaum},
	doi = {10.1103/PhysRevA.73.063626},
	journal = {Phys. Rev. A},
	pages = {063626},
	title = {General variational many-body theory with complete self-consistency for trapped bosonic systems},
	url = {https://doi.org/10.1103/PhysRevA.73.063626},
	volume = {73},
	year = {2006}}

@article{Streltsov:2007,
	author = {Alexej I. Streltsov and Ofir E. Alon and Lorenz S. Cederbaum},
	doi = {10.1103/PhysRevLett.99.030402},
	journal = {Phys. Rev. Lett.},
	pages = {030402},
	title = {Role of Excited States in the Splitting of a Trapped Interacting Bose-Einstein Condensate by a Time-Dependent Barrier},
	url = {https://doi.org/10.1103/PhysRevLett.99.030402},
	volume = {99},
	year = {2007}}

@article{Alon:2007,
	author = {Ofir E. Alon and Alexej I. Streltsov and Lorenz S. Cederbaum},
	doi = {10.1063/1.2771159},
	journal = {J. Chem. Phys.},
	pages = {154103},
	title = {Unified view on multiconfigurational time propagation for systems consisting of identical particles},
	url = {https://doi.org/10.1063/1.2771159},
	volume = {127},
	year = {2007}}

@article{Alon:2008,
	author = {O. E. Alon and A. I. Streltsov and L. S. Cederbaum},
	doi = {10.1103/PhysRevA.77.033613},
	journal = {Phys. Rev. A},
	pages = {033613},
	publisher = {American Physical Society},
	title = {Multiconfigurational time-dependent Hartree method for bosons: Many-body dynamics of bosonic systems},
	url = {https://link.aps.org/doi/10.1103/PhysRevA.77.033613},
	volume = {77},
	year = {2008}}

@article{Lode:2012,
	author = {Axel U. J. Lode and Kaspar Sakmann and Ofir E. Alon and Lorenz S. Cederbaum and Alexej I. Streltsov},
	doi = {10.1103/PhysRevA.86.063606},
	journal = {Phys. Rev. A},
	pages = {063606},
	title = {Numerically exact quantum dynamics of bosons with time-dependent interactions of harmonic type},
	url = {https://doi.org/10.1103/PhysRevA.86.063606},
	volume = {86},
	year = {2012}}

@article{Lode:2016,
	author = {A. U. J. Lode},
	doi = {10.1103/PhysRevA.93.063601},
	journal = {Phys. Rev. A},
	pages = {063601},
	publisher = {American Physical Society},
	title = {Multiconfigurational time-dependent Hartree method for bosons with internal degrees of freedom: Theory and composite fragmentation of multicomponent Bose-Einstein condensates},
	url = {https://link.aps.org/doi/10.1103/PhysRevA.93.063601},
	volume = {93},
	year = {2016}}

@article{Fasshauer:2016,
	author = {Elke Fasshauer and A. U. J. Lode},
	doi = {10.1103/PhysRevA.93.033635},
	journal = {Phys. Rev. A},
	pages = {033635},
	publisher = {American Physical Society},
	title = {Multiconfigurational time-dependent Hartree method for fermions: Implementation, exactness, and few-fermion tunneling to open space},
	url = {https://link.aps.org/doi/10.1103/PhysRevA.93.033635},
	volume = {93},
	year = {2016}}

@article{Lin:2020,
	author = {R. Lin and P. Molignini and L. Papariello and M. C. Tsatsos and C. L{\'e}v{\^e}que and S. E. Weiner and E. Fasshauer and R. Chitra},
	doi = {10.1088/2058-9565/ab788b},
	journal = {Quantum Sci. Technol.},
	pages = {024004},
	title = {MCTDH-X: The multiconfigurational time-dependent Hartree method for indistinguishable particles software},
	url = {https://iopscience.iop.org/article/10.1088/2058-9565/ab788b},
	volume = {5},
	year = {2020}}

@article{Lode:2020,
	author = {A. U. J. Lode and C. L{\'e}v{\^e}que and L. B. Madsen and A. I. Streltsov and O. E. Alon},
	doi = {10.1103/RevModPhys.92.011001},
	journal = {Rev. Mod. Phys},
	pages = {011001},
	publisher = {American Physical Society},
	title = {Colloquium: Multiconfigurational time-dependent Hartree approaches for indistinguishable particles},
	url = {https://journals.aps.org/rmp/abstract/10.1103/RevModPhys.92.011001},
	volume = {92},
	year = {2020}}

@book{TDVM81,
	author = {Kramer, P. and Saraceno, M.},
	publisher = {Springer},
	series = {Lecture Notes in Physics},
	title = {Geometry of the Time-Dependent Variational Principle in Quantum Mechanics},
	volume = {140},
	year = {1981}}

@article{Molignini:2025-SciPost,
	author = {P. Molignini and S. Dutta and E. Fasshauer},
	journal = {SciPost Phys. Lect. Notes},
	volume = {94},
	year = {2025},
	title = {Lecture Notes: many-body quantum dynamics with MCTDH-X},
	}

@misc{MCTDHX,
	author = {A. U. J. Lode and M. C. Tsatsos and E. Fasshauer and S. E. Weiner and R. Lin and L. Papariello and P. Molignini and C. L{\'e}v{\^e}que and M. B{\"u}ttner and J. Xiang and S. Dutta and Y. Bilinskaya},
	title = {MCTDH-X: The MultiConfigurational Time-Dependent Hartree Method for Indistinguishable Particles Software},
	url = {http://ultracold.org},
	year = {2024}}

@article{Xiang:2023,
	author = {Jiabing Xiang and Paolo Molignini and Miriam B{\"u}ttner and Axel U. J. Lode},
	doi = {10.21468/SciPostPhys.14.1.003},
	journal = {SciPost Phys.},
	pages = {003},
	title = {Pauli crystal melting in shaken optical traps},
	url = {https://scipost.org/10.21468/SciPostPhys.14.1.003},
	volume = {14},
	year = {2023}}

@article{Beinke:2018,
	author = {Raphael Beinke and Lorenz S. Cederbaum and Ofir E. Alon},
	doi = {10.1103/PhysRevA.98.053634},
	journal = {Phys. Rev. A},
	pages = {053634},
	title = {Enhanced many-body effects in the excitation spectrum of a weakly interacting rotating Bose-Einstein condensate},
	url = {https://doi.org/10.1103/PhysRevA.98.053634},
	volume = {98},
	year = {2018}}

@article{Roy:2018,
	author = {R. Roy and A. Gammal and M. C. Tsatsos and B. Chatterjee and B. Chakrabarti and A. U. J. Lode},
	doi = {10.1103/PhysRevA.97.043625},
	journal = {Phys. Rev. A},
	pages = {043625},
	title = {Phases, many-body entropy measures, and coherence of interacting bosons in optical lattices},
	url = {https://doi.org/10.1103/PhysRevA.97.043625},
	volume = {97},
	year = {2018}}

@article{Dutta:2019,
	author = {S. Dutta and Marios C Tsatsos and Saurabh Basu and Axel U. J. Lode},
	doi = {10.1088/1367-2630/ab117d},
	journal = {New Journal of Physics},
	pages = {053044},
	title = {Management of the correlations of UltracoldBosons in triple wells},
	url = {https://doi.org/10.1088/1367-2630/ab117d},
	volume = {21},
	year = {2019}}

@article{Schaefer:2020,
	author = {Frank Sch{\"a}fer and Miguel A. Bastarrachea-Magnani and Axel U. J. Lode and Laurent de Forges de Parny and Andreas Buchleitner},
	doi = {10.3390/e22040382},
	journal = {Entropy},
	pages = {382},
	title = {Spectral Structure and Many-Body Dynamics of Ultracold Bosons in a Double-Well},
	url = {https://doi.org/10.3390/e22040382},
	volume = {22},
	year = {2020}}

@article{Lode:2021,
	author = {Axel U. J. Lode and Sunayana Dutta and Camille L{\'e}v{\^e}que},
	doi = {10.3390/e23040392},
	journal = {Entropy},
	pages = {392},
	title = {Dynamics of Ultracold Bosons in Artificial Gauge Fields---Angular Momentum, Fragmentation, and the Variance of Entropy},
	url = {https://doi.org/10.3390/e23040392},
	volume = {23},
	year = {2021}}

@article{Lode:2021-10,
	author = {Axel U. J. Lode and Rui Lin and Miriam B{\"u}ttner and Luca Papariello and Camille L{\'e}v{\^e}que and R. Chitra and Marios C. Tsatsos and Dieter Jaksch and Paolo Molignini},
	doi = {10.1103/PhysRevA.104.L041301},
	journal = {Phys. Rev. A},
	pages = {L041301},
	title = {Optimized observable readout from single-shot images of ultracold atoms via machine learning},
	url = {https://doi.org/10.1103/PhysRevA.104.L041301},
	volume = {104},
	year = {2021}}

@misc{Debnath:2023,
	archiveprefix = {arXiv},
	author = {Pankaj Kumar Debnath and Barnali Chakrabarti and Mantile Leslie Lekala},
	doi = {10.48550/arXiv.2312.13532},
	eprint = {2312.13532},
	title = {Quench dynamics of a Tonks-Girardeau gas in one dimensional anharmonic trap},
	url = {https://doi.org/10.48550/arXiv.2312.13532},
	year = {2023}}

@article{Dutta:2023,
	author = {S. Dutta and A. U. J. Lode and O. Alon},
	doi = {10.1038/s41598-023-29516-w},
	journal = {Sci Rep},
	pages = {3343},
	title = {Fragmentation and correlations in a rotating Bose--Einstein condensate undergoing breakup},
	url = {https://doi.org/10.1038/s41598-023-29516-w},
	volume = {13},
	year = {2023}}

@article{Roy:2023,
	author = {Rhombik Roy and Barnali Chakrabarti and Arnaldo Gammal},
	doi = {10.21468/SciPostPhysCore.6.4.073},
	journal = {SciPost Phys. Core},
	pages = {073},
	title = {Out of equilibrium many-body expansion dynamics of strongly interacting bosons},
	url = {https://doi.org/10.21468/SciPostPhysCore.6.4.073},
	volume = {6},
	year = {2023}}

@article{Dutta:2024,
	author = {Sunayana Dutta and Axel U. J. Lode and Ofir E. Alon},
	doi = {10.1088/1742-6596/2894/1/012014},
	journal = {J. Phys.: Conf. Ser.},
	pages = {012014},
	title = {Condensates breaking up under rotation},
	url = {https://doi.org/10.1088/1742-6596/2894/1/012014},
	volume = {2894},
	year = {2024}}

@article{Haldar:2024,
	author = {Sudip Kumar Haldar and Anal Bhowmik},
	doi = {10.3390/atoms12120066},
	journal = {Atoms},
	pages = {66},
	title = {Many-Body Effects in a Composite Bosonic Josephson Junction},
	url = {https://doi.org/10.3390/atoms12120066},
	volume = {12},
	year = {2024}}

@article{Roy:2024-09,
	author = {Rhombik Roy and Ofir E. Alon},
	doi = {10.1103/PhysRevA.111.043307},
	journal = {Phys. Rev. A},
	pages = {043307},
	title = {Assessing small accelerations using a bosonic Josephson junction},
	url = {https://journals.aps.org/pra/abstract/10.1103/PhysRevA.111.043307},
	volume = {111},
	year = {2025}}

@article{Roy:2024-11,
	author = {Rhombik Roy and Sunayana Dutta and Ofir E. Alon},
	doi = {10.1038/s41598-025-07144-w},
	journal = {Scientific Reports},
	pages = {27193},
	title = {Rotation quenches in trapped bosonic systems},
	url = {https://doi.org/10.1038/s41598-025-07144-w},
	volume = {15},
	year = {2025}}

@article{Bhowmik:2025,
	author = {Anal Bhowmik and Ofir E Alon},
	doi = {10.1088/1367-2630/ada0d3},
	journal = {New Journal of Physics},
	pages = {123035},
	title = {Interference of longitudinal and transversal fragmentations in the Josephson tunneling dynamics of Bose--Einstein condensates},
	url = {https://doi.org/10.1088/1367-2630/ada0d3},
	volume = {26},
	year = {2025}}

@article{Chakrabarti:2025-2,
	journal = {arXiv},
	author = {Barnali Chakrabarti and Arnaldo Gammal},
	doi = {10.48550/arXiv.2502.14440},
	eprint = {2502.14440},
	title = {Localization versus incommemsurability for finite boson system in one-dimensional disordered lattice},
	url = {https://doi.org/10.48550/arXiv.2502.14440},
	year = {2025}}

@article{Fischer:2015,
	author = {Uwe R. Fischer and Axel U. J. Lode and Budhaditya Chatterjee},
	doi = {10.1103/PhysRevA.91.063621},
	journal = {Phys. Rev. A},
	pages = {063621},
	publisher = {American Physical Society},
	title = {Condensate fragmentation as a sensitive measure of the quantum many-body behavior of bosons with long-range interactions},
	url = {https://link.aps.org/doi/10.1103/PhysRevA.91.063621},
	volume = {91},
	year = {2015}}

@article{Chatterjee:2018,
	author = {Chatterjee, Budhaditya and Lode, Axel U. J.},
	doi = {10.1103/PhysRevA.98.053624},
	issue = {5},
	journal = {Phys. Rev. A},
	month = {Nov},
	numpages = {8},
	pages = {053624},
	publisher = {American Physical Society},
	title = {Order parameter and detection for a finite ensemble of crystallized one-dimensional dipolar bosons in optical lattices},
	url = {https://link.aps.org/doi/10.1103/PhysRevA.98.053624},
	volume = {98},
	year = {2018}}

@article{Bera:2019,
	author = {S. Bera and B. Chakrabarti and A. Gammal and M. C. Tsatsos and M. L. Lekala and B. Chatterjee and C. L{\'e}v{\^e}que and A. U. J. Lode},
	doi = {10.1038/s41598-019-53179-1},
	journal = {Scientific Reports},
	pages = {17873},
	title = {Sorting Fermionization from Crystallization in Many-Boson Wavefunctions},
	url = {https://doi.org/10.1038/s41598-019-53179-1},
	volume = {9},
	year = {2019}}

@article{Chatterjee:2019,
	author = {Budhaditya Chatterjee and Marios C Tsatsos and Axel U J Lode},
	doi = {10.1088/1367-2630/aafa93},
	journal = {New Journal of Physics},
	month = {mar},
	number = {3},
	pages = {033030},
	publisher = {{IOP} Publishing},
	title = {Correlations of strongly interacting one-dimensional ultracold dipolar few-boson systems in optical lattices},
	url = {https://doi.org/10.1088/1367-2630/aafa93},
	volume = {21},
	year = 2019}

@article{Bera:2019-symm,
	author = {Sangita Bera and Luca Salasnich and Barnali Chakrabarti},
	doi = {10.3390/sym11070909},
	journal = {Symmetry},
	pages = {909},
	title = {Correlation Dynamics of Dipolar Bosons in 1D Triple Well Optical Lattice},
	url = {https://doi.org/10.3390/sym11070909},
	volume = {11},
	year = {2019}}

@article{Chatterjee:2020,
	author = {Chatterjee, Budhaditya and L\'ev\^eque, Camille and Schmiedmayer, J\"org and Lode, Axel U. J.},
	doi = {10.1103/PhysRevLett.125.093602},
	issue = {9},
	journal = {Phys. Rev. Lett.},
	month = {Aug},
	numpages = {7},
	pages = {093602},
	publisher = {American Physical Society},
	title = {Detecting One-Dimensional Dipolar Bosonic Crystal Orders via Full Distribution Functions},
	url = {https://link.aps.org/doi/10.1103/PhysRevLett.125.093602},
	volume = {125},
	year = {2020}}

@article{Roy:2022,
	author = {Rhombik Roy and Barnali Chakrabarti and Andrea Trombettoni},
	doi = {10.1140/epjd/s10053-022-00345-2},
	journal = {Eur. Phys. J. D},
	pages = {24},
	title = {Quantum dynamics of few dipolar bosons in a double-well potential},
	url = {https://doi.org/10.1140/epjd/s10053-022-00345-2},
	volume = {76},
	year = {2022}}

@article{Hughes:2023,
	author = {Michael Hughes and Axel U. J. Lode and Dieter Jaksch and Paolo Molignini},
	doi = {10.1103/PhysRevA.107.033323},
	journal = {Phys. Rev. A},
	pages = {033323},
	title = {Accuracy of quantum simulators with ultracold dipolar molecules: A quantitative comparison between continuum and lattice descriptions},
	url = {https://doi.org/10.1103/PhysRevA.107.033323},
	volume = {107},
	year = {2023}}

@article{Bilinskaya:2024,
	author = {Yuliya Bilinskaya and Michael Hughes and Paolo Molignini},
	doi = {10.1103/PhysRevResearch.6.L042024},
	journal = {Phys. Rev. Research},
	pages = {L042024},
	title = {Realizing multiband states with ultracold dipolar quantum simulators},
	url = {https://doi.org/10.1103/PhysRevResearch.6.L042024},
	volume = {6},
	year = {2024}}

@article{Molignini:2024,
	author = {P. Molignini and B. Chakrabarti},
	doi = {10.48550/arXiv.2401.10317},
	journal = {arXiv:2401.10317},
	title = {Super-Tonks-Girardeau quench of dipolar bosons in a one-dimensional optical lattice},
	url = {https://doi.org/10.48550/arXiv.2401.10317},
	year = {2024}}

@article{Molignini:2024-2,
	author = {P. Molignini and B. Chakrabarti},
	doi = {10.1088/1367-2630/ad80b8},
	journal = {New J. Phys.},
	pages = {103030},
	title = {Unbounded entropy production and violent fragmentation for repulsive-to-attractive interaction quench in long-range interacting systems},
	url = {https://doi.org/10.1088/1367-2630/ad80b8},
	volume = {26},
	year = {2024}}

@article{Roy:2024-annals,
	author = {Subhrajyoti Roy and Rhombik Roy and Arnaldo Gammal and Barnali Chakrabarti and Budhaditya Chatterjee},
	doi = {10.1016/j.aop.2024.169807},
	journal = {Annals of Physics},
	pages = {169807},
	title = {Quasi-superfluid and Quasi-Mott phases of strongly interacting bosons in shallow optical lattice},
	url = {https://doi.org/10.1016/j.aop.2024.169807},
	volume = {470},
	year = {2024}}

@article{Roy:2024-epjp,
	author = {Rhombik Roy and Andrea Trombettoni and Barnali Chakrabarti},
	doi = {10.1140/epjp/s13360-024-05651-9},
	journal = {Eur. Phys. J. Plus},
	pages = {831},
	title = {Expansion of strongly interacting dipolar bosons in 1D optical lattices},
	url = {https://doi.org/10.1140/epjp/s13360-024-05651-9},
	volume = {139},
	year = {2024}}

@misc{Molignini:2025-quasicryst1,
	archiveprefix = {arXiv},
	author = {Paolo Molignini},
	doi = {10.48550/arXiv.2403.04830},
	eprint = {2403.04830},
	primaryclass = {cond-mat.quant-gas},
	title = {Stability of quasicrystalline ultracold fermions to dipolar interactions},
	url = {https://doi.org/10.48550/arXiv.2403.04830},
	year = {2024}}

@misc{Molignini:2025-quasicryst2,
	author = {Paolo Molignini and Barnali Chakrabarti},
	doi = {10.48550/arXiv.2502.04429},
	eprint = {2502.04429},
	title = {Stability of dipolar bosons in a quasiperiodic potential},
	url = {https://doi.org/10.48550/arXiv.2502.04429},
	year = {2025}}

@misc{Chakrabarti:2025,
	author = {Barnali Chakrabarti and N D Chavda and Andrea Trombettoni and Arnaldo Gammal},
	doi = {10.48550/arXiv.2502.06437},
	eprint = {2502.06437},
	title = {Transport of ultracold dipolar fermions in one-dimensional optical lattices},
	url = {https://doi.org/10.48550/arXiv.2502.06437},
	year = {2025}}

@article{Molignini:2025-JPCM,
	author = {Paolo Molignini},
	doi = {10.1088/1361-648X/ae0fd3},
	journal = {J. Phys.: Condens. Matter},
	pages = {445401},
	title = {Beyond-mean-field phases of rotating dipolar condensates},
	url = {https://doi.org/10.1088/1361-648X/ae0fd3},
	volume = {37},
	year = {2025}}

@article{Lode:2017,
	author = {Lode, Axel U.J. and Bruder, Christoph},
	doi = {10.1103/PhysRevLett.118.013603},
	journal = {Phys. Rev. Lett.},
	pages = {13603},
	title = {Fragmented Superradiance of a Bose-Einstein Condensate in an Optical Cavity},
	url = {https://doi.org/10.1103/PhysRevLett.118.013603},
	volume = {118},
	year = {2017}}

@article{Lode:2018,
	author = {Lode, Axel U.J. and Diorico, Fritz S. and Wu, Rugway and Molignini, Paolo and Papariello, Luca and Lin, Rui and {L{\'{e}}v{\^{e}} Que}, Camille and Exl, Lukas and Tsatsos, Marios C. and Chitra, R. and Mauser, Norbert J.},
	doi = {10.1088/1367-2630/aabc3a},
	journal = {New J. Phys.},
	pages = {055006},
	title = {Many-body physics in two-component Bose-Einstein condensates in a cavity: Fragmented superradiance and polarization},
	url = {https://doi.org/10.1088/1367-2630/aabc3a},
	volume = {20},
	year = {2018}}

@article{Molignini:2018,
	author = {Molignini, Paolo and Papariello, Luca and Lode, Axel U.J. and Chitra, R.},
	doi = {10.1103/PhysRevA.98.053620},
	journal = {Phys. Rev. A},
	pages = {053620},
	title = {Superlattice switching from parametric instabilities in a driven-dissipative Bose-Einstein condensate in a cavity},
	url = {https://doi.org/10.1103/PhysRevA.98.053620},
	volume = {98},
	year = {2018}}

@article{Lin:2019,
	author = {Lin, Rui and Papariello, Luca and Molignini, Paolo and Chitra, R. and Lode, Axel U. J.},
	doi = {10.1103/physreva.100.013611},
	journal = {Phys. Rev. A},
	pages = {013611},
	title = {Superfluid--Mott-insulator transition of ultracold superradiant bosons in a cavity},
	url = {https://doi.org/10.1103/physreva.100.013611},
	volume = {100},
	year = {2019}}

@article{Lin2:2019,
	author = {Lin, Rui and Molignini, Paolo and Lode, Axel U. J. and Chitra, R.},
	doi = {10.1103/PhysRevA.101.061602},
	journal = {Phys. Rev. A},
	pages = {061602(R)},
	title = {Pathway to chaos through hierarchical superfluidity in a cavity-BEC system},
	url = {https://doi.org/10.1103/PhysRevA.101.061602},
	volume = {101},
	year = {2020}}

@article{Lin:2020-PRA,
	author = {Rui Lin and Paolo Molignini and Axel U. J. Lode and R. Chitra},
	doi = {10.1103/PhysRevA.101.061602},
	journal = {Phys. Rev. A},
	pages = {061602(R)},
	title = {Pathway to chaos through hierarchical superfluidity in blue-detuned cavity-BEC systems},
	url = {https://doi.org/10.1103/PhysRevA.101.061602},
	volume = {101},
	year = {2020}}

@article{Lin:2021,
	author = {Rui Lin and Christoph Georges and Jens Klinder and Paolo Molignini and Miriam Buettner and Axel U. J. Lode and R. Chitra and Andreas Hemmerich and Hans Kessler},
	doi = {10.21468/SciPostPhys.11.2.030},
	journal = {SciPost Phys.},
	pages = {030},
	title = {Mott transition in a cavity-boson system: A quantitative comparison between theory and experiment},
	url = {https://doi.org/10.21468/SciPostPhys.11.2.030},
	volume = {11},
	year = {2021}}

@article{Molignini:2022,
	author = {Paolo Molignini and Camille L\'{e}v\^{e}que and Hans Kessler and Dieter Jaksch and R. Chitra and Axel U. J. Lode},
	doi = {10.1103/PhysRevA.106.L011701},
	journal = {Phys. Rev. A},
	pages = {L011701},
	title = {Crystallization via cavity-assisted infinite-range interactions},
	url = {https://doi.org/10.1103/PhysRevA.106.L011701},
	volume = {106},
	year = {2022}}

@article{Rosa-Medina:2022,
	author = {Rodrigo Rosa-Medina and Francesco Ferri and Fabian Finger and Nishant Dogra and Katrin Kroeger and Rui Lin and R. Chitra and Tobias Donner and Tilman Esslinger},
	doi = {10.1103/PhysRevLett.128.143602},
	journal = {Phys. Rev. Lett.},
	pages = {143602},
	title = {Observing Dynamical Currents in a Non-Hermitian Momentum Lattice},
	url = {https://doi.org/10.1103/PhysRevLett.128.143602},
	volume = {128},
	year = {2022}}

@article{Ortuno-Gonzalez:2025,
	author = {Daniel Ortu{\~n}o-Gonzalez and Rui Lin and Justyna Stefaniak and Alexander Baumg{\"a}rtner and Gabriele Natale and Tobias Donner and R. Chitra},
	doi = {https://doi.org/10.48550/arXiv.2505.02837},
	journal = {arXiv:2505.02837},
	title = {Pauli crystal superradiance},
	url = {https://arxiv.org/abs/2505.02837},
	year = {2025}}

@article{Dalfovo:1999,
	author = {F. Dalfovo and S. Giorgini and L. P. Pitaevskii and S. Stringari},
	doi = {10.1103/RevModPhys.71.463},
	journal = {Rev. Mod. Phys.},
	pages = {463},
	title = {Theory of Bose-Einstein condensation in trapped gases},
	url = {https://doi.org/10.1103/RevModPhys.71.463},
	volume = {71},
	year = {1999}}

@book{Jackson:1998,
	address = {New York},
	author = {Jackson, John David},
	edition = {3},
	publisher = {Wiley},
	title = {Classical Electrodynamics},
	year = {1998}}

@article{White:2014,
	author = {White, Andrew C. and Anderson, Brian P. and Bagnato, Vanderlei S.},
	doi = {10.1073/pnas.1312736110},
	journal = {Proceedings of the National Academy of Sciences},
	pages = {4719--4726},
	title = {Vortices and turbulence in trapped atomic condensates},
	volume = {111},
	year = {2014}}

@article{Bradley:2012,
	author = {Bradley, Ashton S. and Anderson, Robert P. and Reichl, Matthew D. and Davis, Matthew J.},
	doi = {10.1103/PhysRevX.2.041001},
	journal = {Physical Review X},
	number = {4},
	pages = {041001},
	title = {Energy Spectra of Vortex Distributions in Two-Dimensional Quantum Turbulence},
	volume = {2},
	year = {2012}}

@article{Chevy:2002,
	author = {Chevy, F. and Bretin, V. and Rosenbusch, P. and Madison, K. W. and Dalibard, J.},
	doi = {10.1103/PhysRevLett.88.250402},
	journal = {Physical Review Letters},
	pages = {250402},
	title = {Transverse Breathing Mode of an Elongated {B}ose-{E}instein Condensate},
	url = {https://doi.org/10.1103/PhysRevLett.88.250402},
	volume = {88},
	year = {2002}}

@article{Bretin:2003,
	author = {Bretin, V. and Rosenbusch, P. and Chevy, F. and Shlyapnikov, G. V. and Dalibard, J.},
	doi = {10.1103/PhysRevLett.90.100403},
	journal = {Physical Review Letters},
	pages = {100403},
	title = {Quadrupole Oscillation of a Single-Vortex Condensate: Evidence for Kelvin Modes},
	url = {https://doi.org/10.1103/PhysRevLett.90.100403},
	volume = {90},
	year = {2003}}

@article{Stenger:1999,
	author = {Stenger, J. and Inouye, S. and Chikkatur, A. P. and Stamper-Kurn, D. M. and Pritchard, D. E. and Ketterle, W.},
	doi = {10.1103/PhysRevLett.82.4569},
	journal = {Physical Review Letters},
	pages = {4569--4573},
	title = {Bragg Spectroscopy of a {B}ose-{E}instein Condensate},
	url = {https://doi.org/10.1103/PhysRevLett.82.4569},
	volume = {82},
	year = {1999}}

@article{Steinhauer:2003,
	author = {Steinhauer, J. and Katz, N. and Ozeri, R. and Davidson, N. and Tozzo, C. and Dalfovo, F.},
	doi = {10.1103/PhysRevLett.90.060404},
	journal = {Physical Review Letters},
	pages = {060404},
	title = {Bragg Spectroscopy of the Multibranch Bogoliubov Spectrum of Elongated {B}ose-{E}instein Condensates},
	url = {https://doi.org/10.1103/PhysRevLett.90.060404},
	volume = {90},
	year = {2003}}

@book{var1,
	address = {Oxford, UK},
	author = {J. Frenkel},
	publisher = {Clarendon Press},
	title = {Wave Mechanics},
	year = {1934}}

@article{var2,
	author = {P. A. M. Dirac},
	doi = {10.1017/S0305004100016108},
	journal = {Proc. Cambridge Philos. Soc.},
	pages = {376},
	title = {Note on Exchange Phenomena in the Thomas Atom},
	url = {https://doi.org/10.1017/S0305004100016108},
	volume = {26},
	year = {1930}}

\end{document}